\documentclass[3p,onecolumn]{elsarticle}
\usepackage{ifpdf}
\usepackage[usenames]{color}
\usepackage{graphicx}
\usepackage{amssymb}
\usepackage{amsmath}
\usepackage{color}
\journal{Communications in Nonlinear Science and Numerical Simulation}

\begin{document}

\begin{frontmatter}

\title{Chimera at the phase-flip transition of an ensemble of identical nonlinear oscillators}

\author[sas]{R.~Gopal}
\author[sas]{V.~K.~Chandrasekar \corref{cor1}}
\ead{chandru25nld@gmail.com}
\author[iis]{D.~V.~Senthilkumar \corref{cor2}}
\ead{skumar@iisertvm.ac.in}
\author[nmc]{A.~Venkatesan}
\author[cnld]{M.~Lakshmanan}

\cortext[cor1]{Corresponding author} 
\cortext[cor2]{Corresponding author} 

\address [sas]{Centre for Nonlinear Science \& Engineering, School of Electrical \& Electronics Engineering, SASTRA University, Thanjavur- 613 401, India.} 
\address[cnld]{Centre for Nonlinear Dynamics, School of Physics, Bharathidasan University, Tiruchirapalli-620024, India}
\address[iis]{School of Physics, Indian Institute of Science Education and Research, Thiruvananthapuram -695016, India}
\address[nmc]{Department of Physics, Nehru Memorial College, Puthanampatti, Tiruchirapalli 621 007, India}

\date{\today}
   
\begin{abstract}
A complex collective emerging behavior characterized by coexisting coherent and incoherent domains is termed as a chimera state.  We bring out the existence of a new type of chimera in a nonlocally coupled ensemble of identical oscillators driven by a common dynamic environment. The latter facilitates the onset of phase-flip bifurcation/transitions among the coupled oscillators of the ensemble, while the nonlocal coupling induces a partial asynchronization among the out-of-phase synchronized oscillators at this onset. This leads to the  manifestation of coexisting out-of-phase synchronized coherent domains interspersed by asynchronous incoherent domains elucidating the existence of a different type of chimera state.  In addition to this, a rich variety of other collective behaviors such as   clusters with phase-flip transition, conventional chimera, solitary state and complete synchronized state which have been reported using different coupling architectures are  found to be induced by the employed couplings for appropriate coupling strengths. The robustness of the resulting dynamics is demonstrated in ensembles of two paradigmatic models, namely R\"ossler oscillators and Stuart-Landau oscillators.
\end{abstract}

\end{frontmatter}

\section{Introduction}

An ensemble of coupled oscillators is a veritable black box exhibiting a plethora of
complex cooperative dynamical behaviors mimicking several real world phenomena~\cite{Kuramoto:84,Pikovsky:01,boccaletti,Winfree:01}.
Chimera state is such an intriguing emerging behavior that has been
identified in an ensemble of identical oscillators with non-local coupling~\cite{Kuramoto:02,Abrams:04,Abrams:041,Abrams:042,Abrams:043,Laing,Shima:04,Bor:2010,Martens:2010,Sheeba:2009,shepelev2017}.
Since its identification, the notion of chimera  has provoked a flurry of intense investigations
because of the surprising fact that such an ensemble
splits into two dynamically  distinct domains, wherein  all the oscillators evolve in synchrony
in the coherent domain, while the oscillators in the incoherent domain evolve in asynchrony~\cite{Omel:11,Omel:111,Panag:15}. 
Earlier investigations on the phenomenon of chimera were restricted to 
nonlocal coupling, both in the weak and the strong coupling limits, giving rise to
frequency ~\cite{Omel:13,Gopal:15} and amplitude chimeras~\cite{Dud:14,zak:14,Gopal:15,omel12015,omel22015,sem2015,bogomolov}, as it was believed
that nonlocal coupling is a prerequisite for the  onset of chimera in an ensemble of
identical oscillators.  Later investigations revealed the emergence of chimera 
states under global coupling~\cite{bay:14,Sethia:14,Sch:14,Prema:15}, 
mean-field coupling~\cite{baner:14} and  even in nearest neighbor coupling~\cite{Laing:15,Bera:15}.
In addition to the frequency and amplitude chimeras, other types of chimera states 
such as amplitude mediated chimeras~\cite{Sethia:13}, intensity induced chimeras~\cite{Gopal1:14}, and
chimera death~\cite{zak:14,Prema:15,prema:2016,Dutta:15,Sch:15} were also reported in the  recent literature. 
 Further, the robustness of chimera states was reported 
in ~\cite{omel12015,loos2016}, while chimera states and multi-cluster chimera 
 states were also found in oscillators with time delayed feedback~\cite{larger2015,semenov2016,scholl:2016}. 
Very recently, different types of chimera states including
imperfect chimera states and solitary states were also reported at the transition 
from incoherence to coherence~\cite{jaros2015,nade2017}.

\begin{figure}[ht!]
\centering
\includegraphics[width=1.0\linewidth]{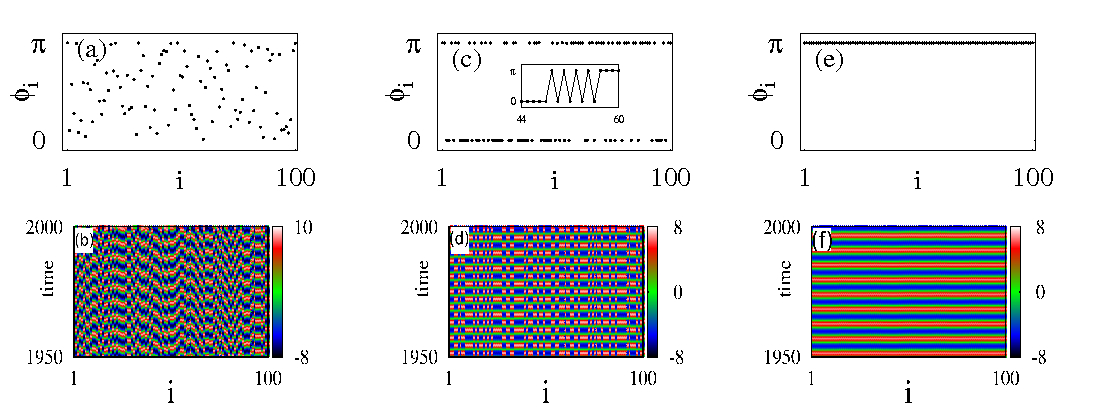}
\caption{(Color online) Snapshots of instantaneous phases $\displaystyle \phi_i=\arctan(y_i/x_i)$ (left column) and  space-time plots of $x_{i}$ (right column) of the R\"ossler oscillators (1) under the influence of the dynamic environment  alone, with no nonlocal interaction(that is $\epsilon=0$).
The parameters are fixed at $a=0.165$, $b=0.4$, $c=8.5$, $k=25$  and $\eta=2$ for different values $q$  
(a)-(b) $q=0.02$, (c)-(d) $q=0.4$, and (e)-(f) $q=1$.  The inset in Fig.~1(c) represents  a two cluster state with phase-flip transition  which is a special case of spatial chaos.}
\label{chaos_fig1}
\end{figure}

 Notably, chimera states have also been  demonstrated in laboratory experiments.
In particular, chimera states have been discovered in populations of coupled  
chemical oscillators, in electro-optical systems~\cite{tinsley:12,hagers:12},
electronic circuits~\cite{Nkoma:13} and in mechanical systems with
two sub-populations of identical metronomes~\cite{martens:13}.
Surprisingly spatiotemporal patterns  mimicking chimera states were also found in
real world systems, which include the unihemispheric sleep of animals~\cite{rottenberg:00}, 
the multiple time scales of sleep dynamics~\cite{Olb:11},  and so on. 
 Very recently, chimera states have also been reported in a network of two
populations of Kuramoto model with inertia~\cite{olmi,olmi1}, which is also  a model used in
the analysis of power grids~\cite{Filatrella:2008}.
Further investigations on identifying the intricacies involved
in the mechanism of the onset of chimera states is of vital importance from the perspective 
of neuroscience because of the concept of ``bumps" of neuronal activity~\cite{laing:2000,harris}
associated with it.

\begin{figure}
\centering
\includegraphics[width=1.0\columnwidth]{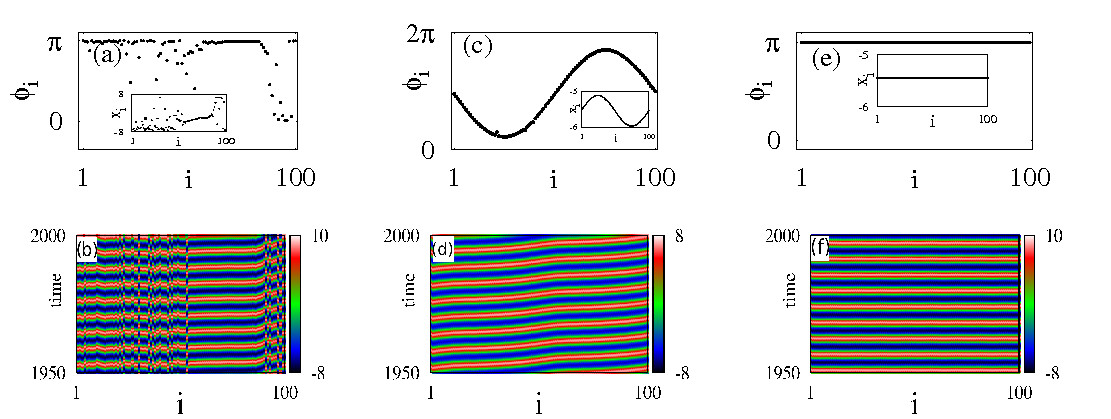}
\caption{(Color online) Snapshots of the instantaneous phases $\phi_{i}$ (left column) (of $x_i$ are shown in the inset) and the space-time evolution of $x_{i}$
 (right column) of the ensemble of R\"ossler oscillators with $q=0.02$ for the coupling radius $r=0.3$ of the 
nonlocal coupling and for different values of the strength of the nonlocal coupling (a)-(b) $\varepsilon=0.01$,
(c)-(d) $\varepsilon=0.02$ and (e)-(f) $\varepsilon=0.05$. Note that the insets in (b) and (c) correspond to the values of the variables $x_{i}$.
The values of the other parameter are the same as in Fig.~\ref{chaos_fig1}(a).  Note that the insets depict the snapshots of the amplitudes $x_{i}$.}
\label{chaos_fig2}
\end{figure}

In this paper, we report an interesting type of chimera arising out of the
phase-flip bifurcation~\cite{Prasad:05,Prasad:051,Prasad:052,McMillen:02} of an ensemble of identical oscillators. The coherent
domains of the chimera state are out-of-phase synchronized  with each other, whereas the incoherent
 domain is constituted by the asynchronous oscillators. 
Both the out-of-phase synchronized coherent  domains and the incoherent domain
coexist simultaneously in an ensemble of identical oscillators for suitable
parameter values attributing to the existence of  a new type of interesting chimera state, namely chimera at the phase-flip bifurcation/transition of the ensemble of oscillators.  
In addition to the above, clusters with phase-flip transition $(PFC)$, conventional
chimera, solitary state and complete synchronized states are also found to 
exist from the completely incoherent oscillator ensemble as a function of the coupling strength.   We find that the phase-flip transition
among the oscillators in the ensemble is induced by a common environmental coupling~\cite{Amit:12},
while the chimera state in the ensemble is induced by the nonlocal coupling among the
identical oscillators.  The phenomenon of phase-flip bifurcation/transition was
shown to be induced by a common environmental coupling in two coupled oscillators
in the recent past~\cite{Amit:12}.  Abrupt transition from in-phase to anti phase synchronized oscillations among the 
oscillators as a function of a system parameter, where their relative
phase difference changes from zero to $\pi$ at the bifurcation/transition point, is known as phase-flip bifurcation/transition~\cite{Prasad:05,Prasad:051,Prasad:052}. 
The indirect coupling arising from the common medium
or from the environment of the dynamical systems was shown to be a source of several 
collective behaviors of real world systems (see Ref.~\cite{Ambi:10,Amit:12,sharma:2012} and references therein). 

\begin{figure}
\centering
\includegraphics[width=1.0\columnwidth]{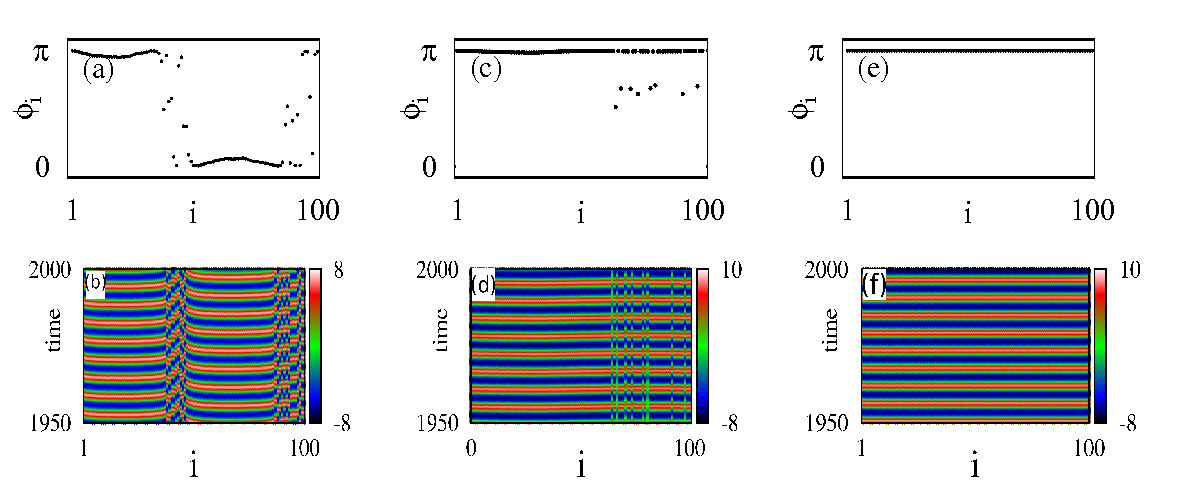}
\caption{(Color online) Snapshots of the instantaneous phases $\phi_{i}$ (left column) and the space-time evolution (right column) of the ensemble of R\"ossler oscillators  with $q=0.4$ for the coupling radius $r=0.3$ of the nonlocal coupling and for different values of the strength of the nonlocal coupling (a)-(b) $\varepsilon=0.03$,
(c)-(d) $\varepsilon=0.051$ and (e)-(f) $\varepsilon=0.1$. 
The values of the other parameter are the same as in Fig.~\ref{chaos_fig1}(c).}
\label{chaos_fig3}
\end{figure}

For instance, the indirect environmental coupling plays a crucial role in facilitating 
complex collective dynamics such as decoherence,  co-ordinated rhythms in biological systems~\cite{harris} and 
quorum  sensing~\cite{camilli2006,bing2012,chen:11,gua:2013}.
Our studies show  that such a type of coupling  along with a nonlocal coupling leads to a rich variety of complex collective behaviour like different types of chimeras, solitary state, and complete synchronized state reported in the literature using different coupling architectures. These are 
necessary and important complements to the current knowledge on the collective behaviors due to the dynamic environmental coupling. In addition,
such a coupling facilitates a new type of interesting chimera, namely chimera at the phase-flip bifurcation/transition of the ensemble of oscillators for appropriate
coupling strengths.

The plan of the paper is as follows.  In Sec.~II, the emergence of chimera at 
the phase-flip transition and the conventional chimera from the
nonlocally coupled R\"ossler oscillators with a common dynamic environment will be discussed.  
Similarly, in Sec.~III, we corroborate the generic nature of the results in the 
nonlocally coupled Stuart-Landau oscillators with a common dynamic environment.  Finally,
in Sec.~IV, we provide the summary and conclusion. We present the existence of  clusters with phase-flip transition, which is a  special case of classical spatial chaos for different sets of initial conditions in Appendix-A, the results of phase-flip transition between two coupled oscillators  in Appendix-B and also discuss the emergence of bistability among them in Appendix-C.

\section{Nonlocally coupled R\"ossler oscillators with common dynamic environment} 
We consider an ensemble of nonlocally coupled identical chaotic R\"ossler oscillators interacting through a 
common dynamic environment  represented as
\begin{subequations}
\begin{align}
\dot{x}_{i}=&\,-y_{i}-z_{i}+\frac{\varepsilon}{2P}\sum_{j=i-P}^{j=i+P}(x_{j}-x_{i}),  \\
\dot{y}_{i}=&\,x_{i}+ay_{i}, \\
\dot{z}_{i}=&\,b+z_{i}(x_{i}-c)+kw_{i}, \\ 
\dot{w}_{i}=&\,-\alpha w_{i}+0.5z_{i}-\eta(w_{i}-\frac{q}{N}\sum_{j=1}^{N}w_{j}), \\
& i=1,2,....,N \nonumber
\end{align}
\label{ross}
\end{subequations}
where $\alpha=1$,$a=0.165$, $b=0.4$ and $c=8.5$  are the system parameters 
and the number of oscillators is fixed as $N=100$  throughout the manuscript.  The evolution of every
individual oscillator is governed by a direct and an indirect coupling, namely
a nonlocal coupling and an environmental coupling, respectively.  $P\in(1,N/2)$ quantifies the number of 
nearest neighbors of any oscillator in the ring with a coupling radius 
$r=\frac{P}{N}$, which provides the measure of nonlocal coupling. $\varepsilon$
is the coupling strength of the nonlocal coupling. We have employed
the  same environmental coupling used to induce phase-flip transition between two coupled
dynamical systems in Ref.~\cite{Amit:12}
 In this coupling, there are agents interacting
with each other in a common dynamic environment and the $i$th individual oscillator in 
the ensemble interacts directly  with the $i$th agent of the environment with a coupling strength $k$. 
Specifically, the evolution equation for each agent $w_i$ is given by Eq.~f(\ref{ross}d), where $\eta$ is a 
diffusion constant and $q$ is the strength of the mean field interaction of all the agents. 
 Since the effect of environmental factors on the oscillators are relatively feeble compared to the interactions 
among the oscillators,  the agents from the common environment usually interact strongly with large coupling strengths
in order to have an appreciable influence on the oscillators.  Hence, the  coupling strength $k$ is usually large as
adopted in Ref.~\cite{Amit:12} to realize the phase-flip bifurcations.
Such environmental couplings play a vital role in biochemical reactions
at the cellular level~\cite{Amit:12,Taylor:09,chen:11,Ullner:07}. The dynamic environmental coupling facilitates the 
onset of the phase-flip
transition among the oscillator ensemble even in the absence of the nonlocal coupling,
whereas the nonlocal coupling induces asynchronous state at the phase-flip transition among the
out-of-phase synchronized coherent domains. Consequently, one is lead to the onset of
coexisting coherent and incoherent domains corroborating  the emergence of
an interesting  new type of chimera state, namely chimera at the phase-flip transition.

In order to realize the above phenomenon, first we will examine the emerging dynamics 
of the ensemble of R\"ossler oscillators (\ref{ross}) 
due to environmental coupling as a function of the strength of the mean field interaction $q$.
 For this the strength of the nonlocal diffusive coupling $\varepsilon$ is fixed as $\varepsilon=0.0$. 
The coupling strength between the agents and the oscillators is fixed as $k=25$,
while the diffusion constant $\eta$ as $\eta=2.0$,  as in Ref.~\cite{Amit:12}, throughout this section.  
 We have used the Runge-Kutta fourth order integration scheme with an integration time step of $0.01$ to solve the dynamical equations.  One can also
 use smaller values of integration time step but  which we find only increases the transient times to observe the stable dynamical states and do not 
 alter our results. Therefore we fix the time step as 0.01 in our entire analysis. In our analysis, we have 
discarded $4\times 10^5$ time steps as transients to ensure that the observed dynamical behaviors are steady state dynamics and not transients. Snapshots of the instantaneous
phases of all the oscillators in the ensemble along with their spatio-temporal evolution are
shown in Fig.~\ref{chaos_fig1} for three different values of the strength of the mean
field coupling $q$. Random initial conditions uniformly distributed between $-1$ to $1$  are
chosen among the ensemble of R\"ossler oscillators. 
The oscillators in the ensemble display an asynchronous state both in the
snapshot of their instantaneous phases (see Fig.~\ref{chaos_fig1}(a)) and in their space-time
plot (see Fig.~\ref{chaos_fig1}(b)) for  $q=0.02$.   Upon increasing the strength of the
mean field coupling, the oscillators get randomly segregated into an in-phase and anti-phase synchronized state
as shown in Figs.~\ref{chaos_fig1}(c) and \ref{chaos_fig1}(d) for $q=0.4$, displaying the
existence of $PFC$, which is a special case of classical spatial chaos~\cite{Omel:11,Omel:111,nizhmik2002}. 
(We also point out in Appendix-A that clusters with phase-flip transition  arise not essentially due to specific choice of initial conditions but more because of environmental coupling by  illustrating that these states emerge for widely different initial conditions).It is also to be noted that even though Figs. \ref{chaos_fig1}(c)  has some resemblance to  Fig. 2 (e)  in Ref.~\cite{Omel:11}, both are dynamically different.  Two coherent regimes which are split into simply a lower branch and an upper branch interspersed by an incoherent regime is depicted in  Fig. 2(e)  of Ref.~\cite{Omel:11}.  In contrast, our results in Fig. ~\ref{chaos_fig1}(c) display not simply two coherent regimes splitting into lower and upper branches but two out-of-phase synchronized coherent regimes (branches)  with a relative phase difference of $\pi$, resembling phase-flip bifurcation/transition in an ensemble of oscillators, interspersed by an incoherent regime. For further larger value of $q$, all the oscillators evolve in synchrony as depicted in Figs.~\ref{chaos_fig1}(e) and ~\ref{chaos_fig1}(f) for $q=1.0$.  

\begin{figure}
\centering
\includegraphics[width=0.9\columnwidth]{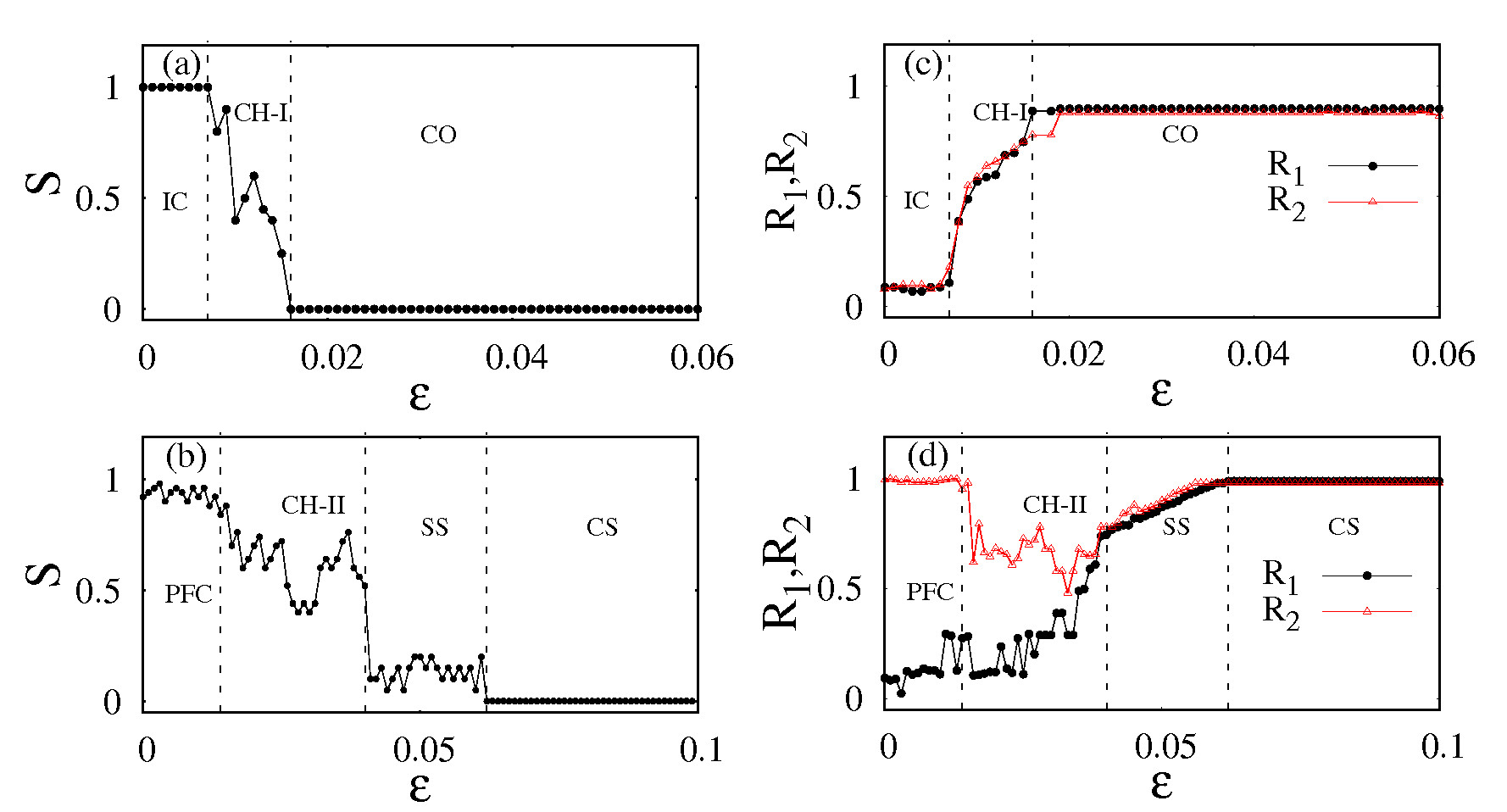}
\caption{(Color online){ The strength of incoherence $S$ ((a)-(b)) and Kuramoto order parameter $R_{1},R_{2}$ ((c)-(d))} as a function of the nonlocal coupling strength $\varepsilon$ for (a),(c) $q=0.02$ and (b),(d) $q=0.4$ characterizing the dynamical transitions
in Fig.~\ref{chaos_fig2} and Fig.~\ref{chaos_fig3}, respectively.}
\label{chaos_fig4}
\end{figure}
\begin{figure}
\centering
\includegraphics[width=0.9\columnwidth]{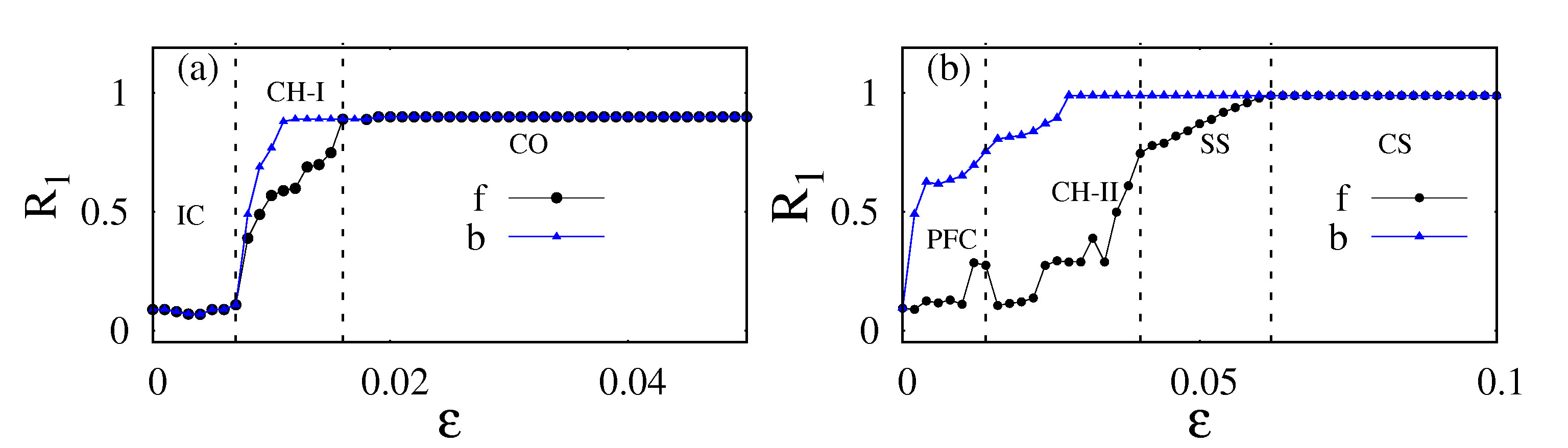}
\caption{(Color online){Kuramoto order parameter $R_{1}$ for both the forward (represented by the line `f' connected by filled circles) and the backward  (represented by the line `b' connected by filled triangles) scanning of the nonlocal coupling strength $\varepsilon$ for (a) $q=0.02$ and (b) $q=0.4$ depicting hysteresis in their dynamical transitions.}
}
\label{hysteresis}
\end{figure}

Now we will unravel the effect of the nonlocal coupling on the asynchronous state and 
the $PFC$ observed for the strengths of the mean field coupling  $q=0.02$ and $q=0.4$ in Figs.~\ref{chaos_fig1}(a, b) and \ref{chaos_fig1}(c, d) respectively.  To start with, we will consider the asynchronous state for $q=0.02$ depicted in Figs.~\ref{chaos_fig1}(a)-\ref{chaos_fig1}(b) 
and increase the strength of the nonlocal coupling $\varepsilon$.   Snapshots of the instantaneous
phases of all the oscillators  along with their spatio-temporal evolution are illustrated in
Figs.~\ref{chaos_fig2} for different values of $\varepsilon$. The radius of the nonlocal coupling
is fixed  as $r=0.3$.  Even for a very low value of the strength of the nonlocal coupling,
the  oscillators  in the ensemble splits into coherent and  incoherent domains as
shown in Figs.~\ref{chaos_fig2}(a) and ~\ref{chaos_fig2}(b) for $\varepsilon=0.01$ elucidating the
existence of conventional chimera $(CH$-$I$). 
It is to be noted that the phases of the  asynchronous oscillators for $\varepsilon=0$ are distributed between
$0$ and $\pi$ (see Fig.~\ref{chaos_fig1}(a)), whereas the phases of the oscillators in the coherent and incoherent domains
comprising the chimera state for a finite value of $\varepsilon$ is shifted towards $\pi$ (see Fig.~\ref{chaos_fig2}(a))
indicating the tendency towards an entertainment of their phases. 
The ensemble of oscillators evolve in coherence ($CO$) as depicted in the snapshot and 
space-time plots in Figs.~\ref{chaos_fig2}(c) and ~\ref{chaos_fig2}(d), respectively, for
$\varepsilon=0.02$, where the coherent evolution of the oscillators is evident from the
inset of Fig.~\ref{chaos_fig2}(c). The oscillators evolve in complete synchrony ($CS$) 
(see the inset of Fig.~\ref{chaos_fig2}(e)) for a further larger value of the 
strength of the nonlocal coupling as shown in  Figs.~\ref{chaos_fig2}(e) and ~\ref{chaos_fig2}(f)
for $\varepsilon=0.05$.

Next, we will illustrate the effect of the nonlocal coupling on the  $PFC$ state
in Fig.~\ref{chaos_fig1}(c)-\ref{chaos_fig1}(d) induced by the mean field coupling with $q=0.4$.
Snapshots of the instantaneous phases of all the oscillators  along with their spatio-temporal 
evolution are illustrated in Figs.~\ref{chaos_fig3} for different values of $\varepsilon$. Now, 
the oscillators  are clearly segregated into two coherent domains exhibiting out-of-phase 
synchronized oscillations  with each other interspersed by a domain of asynchronous oscillators at the onset of 
the phase-flip transition among the oscillators as shown in Fig.~\ref{chaos_fig3}(a) for $\varepsilon=0.03$. 
Thus the coexisting out-of-phase synchronized coherent  domains and the incoherent domain in between the
coherent domains (see also Fig~\ref{chaos_fig3}(b)) constitute  an interesting new type of 
chimera state, namely, chimera  at the phase-flip transition ($CH$-$II$).  
It is to be noted that Figs. ~\ref{chaos_fig3}(a, b) have  close resemblance to  Fig. 2 (d)  of Ref.~\cite{Omel:11} and
Fig. 4(b) of Ref.~\cite{Omel:111}.

Further, we have also identified the existence of imperfectly synchronized states  illustrated in 
Figs.~\ref{chaos_fig3}(c)-\ref{chaos_fig3}(d)  for $\varepsilon=0.051$. The imperfect synchronized states 
are also called solitary states, which refer to excursions of a small number of oscillators 
away from synchronized group~\cite{jaros2015,kapi,nade2017,prema:2016,brez,ym2014}.
Thus a finite nonlocal coupling is necessary to induce the chimera at the phase-flip transition
and the solitary state in addition to the common environmental coupling.
It is clear that either the common environmental coupling or the nonlocal coupling alone
cannot give  rise to these states.
Finally, the ensemble of oscillators evolve synchronously for further increase in the 
strength of the nonlocal coupling as shown in Figs.~\ref{chaos_fig3}(e)-\ref{chaos_fig3}(f) for $\varepsilon=0.1$.

\begin{figure}
\centering
\includegraphics[width=0.7\columnwidth]{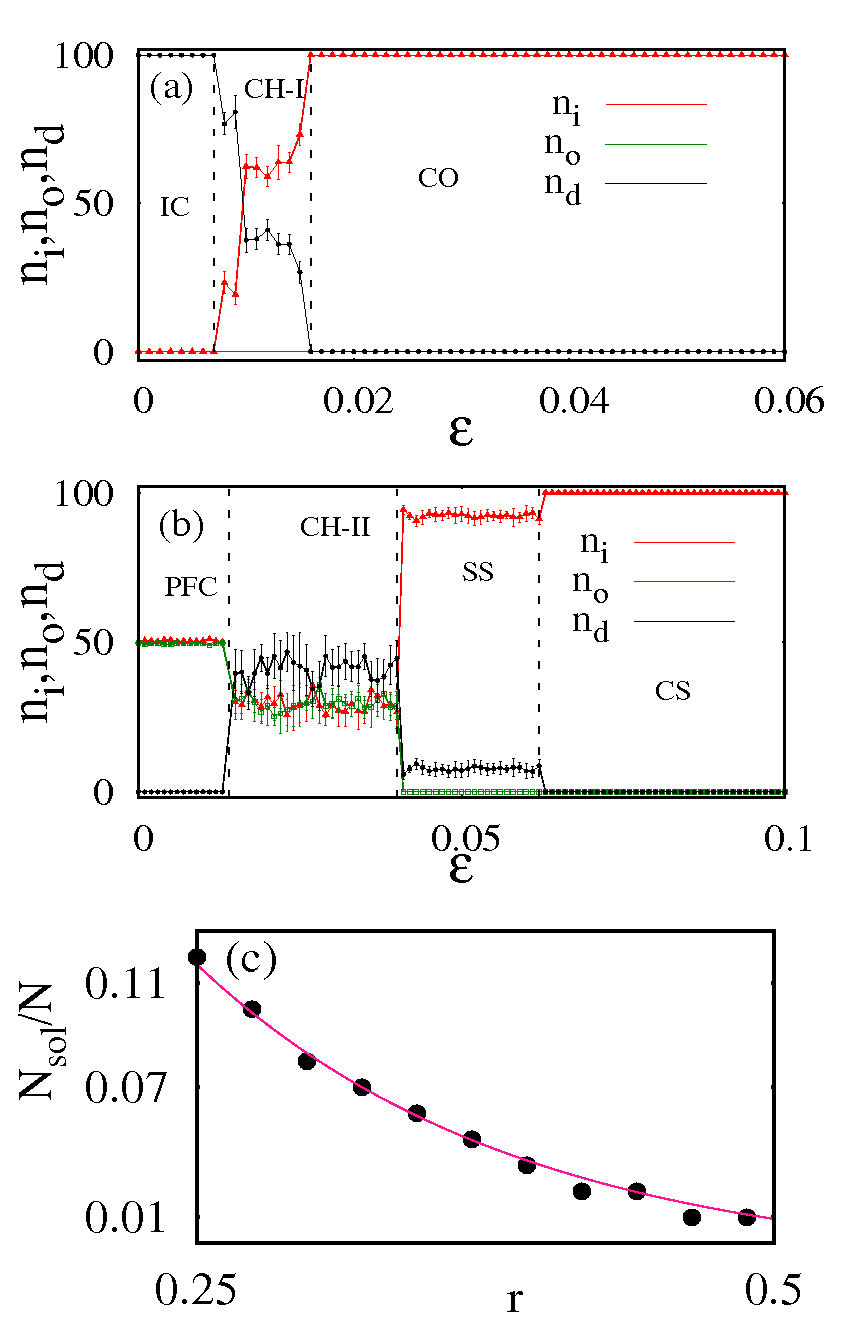}
\caption{(Color online) Number of oscillators exhibiting in-phase oscillations $n_i$,
out-of-phase oscillations $n_o$ and desynchronized state $n_d$  as a function of the
strength of the nonlocal coupling for different values of the coupling radius 
(a) $q=0.02$ and (b) $q=0.4$ characterizing the dynamical transitions
in Fig.~\ref{chaos_fig2} and Fig.~\ref{chaos_fig3}, respectively,  with error bars indicated.  (c) The solitary fraction, namely the ratio of the number of 
desynchronized oscillators (solitary oscillators) $N_{sol}$ to the total number of the oscillators $N$ as a function of the coupling 
radius $r$, displays an exponential decay.}
\label{chaos5}
\end{figure}
\begin{figure}
\centering
\includegraphics[width=0.7\columnwidth]{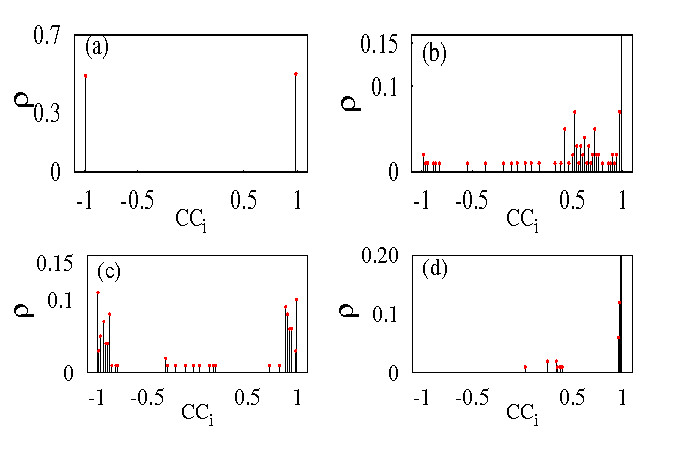}
\caption{(Color online) Probability distribution of correlation  coefficient for different 
collective dynamical states in the ensembles of R\"ossler oscillators: (a) phase flip chimera ($\epsilon=0.0, q=0.4$),
(b) chimera I ($\epsilon=0.01, q=0.02$),
(c) chimera II ($\epsilon=0.03,q=0.4$),
(d) solitary states ($\epsilon=0.051, q=0.4$).}
\label{chaos_fig5_1}
\end{figure}
\subsection{Quantification measures to characterize the chimera states}
Recently a quantitative measure, namely,  the strength of incoherence $S$~\cite{rgvkc2014} was introduced,
to distinguish various collective dynamical states, as
\begin{equation} 
S=1-\frac{\sum_{m=1}^{M}s_{m}}{M}, \hspace{0.1cm}  s_{m}=\Theta(\delta-\sigma_{l}(m)),
\end{equation}
where $\Theta(\cdot)$ is the Heaviside step function, and $\delta$ is a predefined 
threshold chosen as a certain percentage value of the difference between
the upper/lower bounds, $x_{l,i_{max}}/x_{l,i_{min}}$.
$M$ is the number of bins of equal size $n=N/M$.
The local standard deviation $\sigma_{l}(m)$ is introduced as
\begin{equation}
\sigma_{l}(m)=\left<\noindent \sqrt{\frac{1}{n}\sum_{j=n(m-1)+1}^{mn}[z_{l,j}-<z_{l,m}>]^2} 
\hspace{0.1cm}\right>_{t},\\
\end{equation}
$m=1,2,...M,$ \\
where $z_{l,i}=x_{l,i}-x_{l,i+1}$, $l=1,2...d$, $d$ is the dimension of the individual unit in
the ensemble, $i=1,2...N$, $<z_{l,m}>=\frac{1}{n}\sum_{j=n(m-1)+1}^{mn}z_{l,j}(t)$, and
$\langle \cdot\rangle_{t}$ denotes the time average.
When $\sigma_{l}(m)$ is less than $\delta$, $s_{m}=1$, otherwise $s_{m}=0$. 

\begin{figure}
\centering
\includegraphics[width=0.7\linewidth]{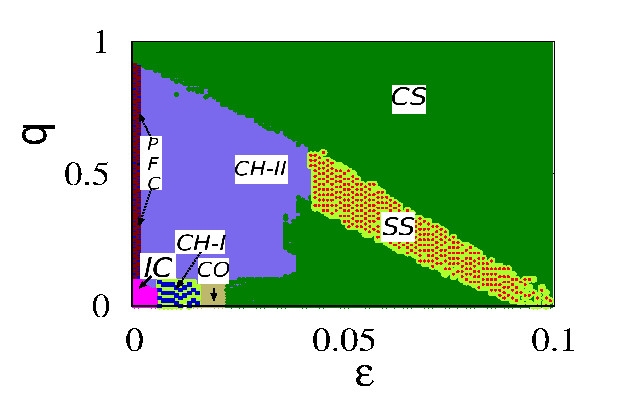}
\caption{(Color online) Two parameter phase diagram depicting the collective dynamical
states of nonlocally coupled R\"ossler oscillators with a common dynamic environment as
a function of the strength of the nonlocal coupling $\varepsilon\in(0,0.1)$ and 
the strength of the mean field coupling $q\in(0,1)$. The parameter regions marked as
$IC$, $PFC$, $CH$-$I$, $CH$-$II$, $SS$, $CO$, and $CS$ correspond to the desynchronized state,
  $PFC$, conventional chimera, chimera at the phase-flip transition, solitary state, 
phase synchronized state and complete synchronized state, respectively.}
\label{chaos_fig6}
\end{figure}
 In the incoherent domain, the local standard deviation $\sigma_{l}(m)$ has 
some finite value  greater than $\delta$ and hence  $s_{m}=0, \forall~m$. Hence,
the strength of incoherence $S$ acquire unit value for the incoherent domain.
On the other hand, in the coherent domain, the
standard deviation $\sigma_{l}(m)$ is always zero  and 
hence $s_{m}=1, \forall~m$,  resulting in the null value of $S$ ($m$ in the present case is chosen as $m=20$).
The strength of incoherence $S$ will have intermediate values between zero and one, $0 < S < 1$
for the chimera states as they are characterized by coexisting coherent and incoherent domains.
The strength of the incoherence $S$  is depicted in Fig.~\ref{chaos_fig4} for the 
mean field interactions $q$ corresponding to Figs.~\ref{chaos_fig2} and~\ref{chaos_fig3} as a function
of the nonlocal coupling $\varepsilon$. The unit value of $S$ 
in the range of $\varepsilon\in(0,0.008)$ in Fig.~\ref{chaos_fig4}(a) for $q=0.02$
indicates that the coupled oscillators evolve in asynchrony, whereas the intermediate
values between zero and unity for $\varepsilon\in(0.008,0.016)$ corroborate  the coexistence
of both coherent and incoherent domains characterizing the existence of chimera states.
For $\varepsilon>0.016$, the null value of the strength of the incoherence confirms
the synchronous evolution of the coupled oscillators. Upon increasing the strength of 
the mean field interaction to $q=0.4$, the spread of the chimera
state is extended  to a larger range of $\varepsilon\in(0.016,0.041)$ as confirmed by the
intermediate values of $S$ in Fig.~\ref{chaos_fig4}(b). Chimera state
is preceded by  $PFC$  in the range of $\varepsilon\in(0.0,0.016)$
 as indicated by the value of $S$ close to unity as there are out-of-phase synchronized
and asynchronous oscillators in all the $m$ domains. Further, the value of $S$
close to zero in the range of $\varepsilon\in(0.041,0.061)$ characterizes the existence of
solitary state with a few asynchronous oscillators in the bins.  Synchronous state 
for $\varepsilon>0.061$ is shown by the  null values of $S$ in Fig.~\ref{chaos_fig4}(b).
In order to confirm the above states further, we also estimated the Kuramoto order parameters  $R_{1}$, $R_{2}$ defined by~\cite{strogatz2000,restrepo2006} 

\begin{equation}
R_{n}=\Bigg\langle \frac{1}{N}\left|\sum_{j=1}^{N} e^{in\theta_{j}}\right|\Bigg \rangle_{t},   ~~~~~~\mbox{where}~n=1,2, 
\label{order}
\end{equation}
Note that here $R_1$ and $R_2$ correspond to the order parameters in Eq.(\ref{order}). Here $R_{1}$ corresponds to in-phase synchronization and $R_{2}$ corresponds to out-of-phase synchronization. 
In particular, values of both $R_1$ and $R_2$  are required to characterize the $PFC$. $R_1$ will acquire 
null value for out-of-phase synchronized oscillators whereas $R_2$ will acquire unit value for the same set of out-of-phase synchronized oscillators
as $R_2=e^{2i\theta}$.  The coupled system exhibits  an
incoherent state when the oscillator phases are uniformly distributed so that $R_{1}\approx R_{2} \approx 0$ 
 and coherent state when their phases are entrained, $R_{1}\approx R_{2} \approx 1$.  
 The order parameter takes intermediate values,  $0\leq R_{1},R_{2}\leq 1$, for chimera state.
The order parameters acquire values $0$ and $1$ for $PFC$ as in the case of two clusters which are out-of-phase with each other (See. Figs ~\ref{chaos_fig4}(d)). 
The Kuramoto order parameter,  corresponding to Figs.~\ref{chaos_fig4}(a-b), depicted in Figs.~\ref{chaos_fig4}(c-d) again confirm
the transitions shown by the strength of incoherence.
The strength of incoherence and the Kuramoto order parameters  in Figs. \ref{chaos_fig4}(a-b) and \ref{chaos_fig4}(c-d), respectively, elucidate the transitions among the  different states are second order transitions as there is a smooth transition from one state to the other. 
 In addition, the  order parameter $R_{1}$ is also depicted  for both the forward scanning (represented by the line `f' connected by
filled circles) and the backward scanning  (represented by the line `b' connected by filled triangles), 
starting from the same final conditions at the end of forward scanning, of the coupling strength  $\epsilon$  in
Figs.~\ref{hysteresis}(a)-(b) for $q=0.02$ and $0.4$, respectively,  illustrating the existence of hysteresis in their dynamical transitions.

\begin{figure}
\centering
\includegraphics[width=0.7\columnwidth]{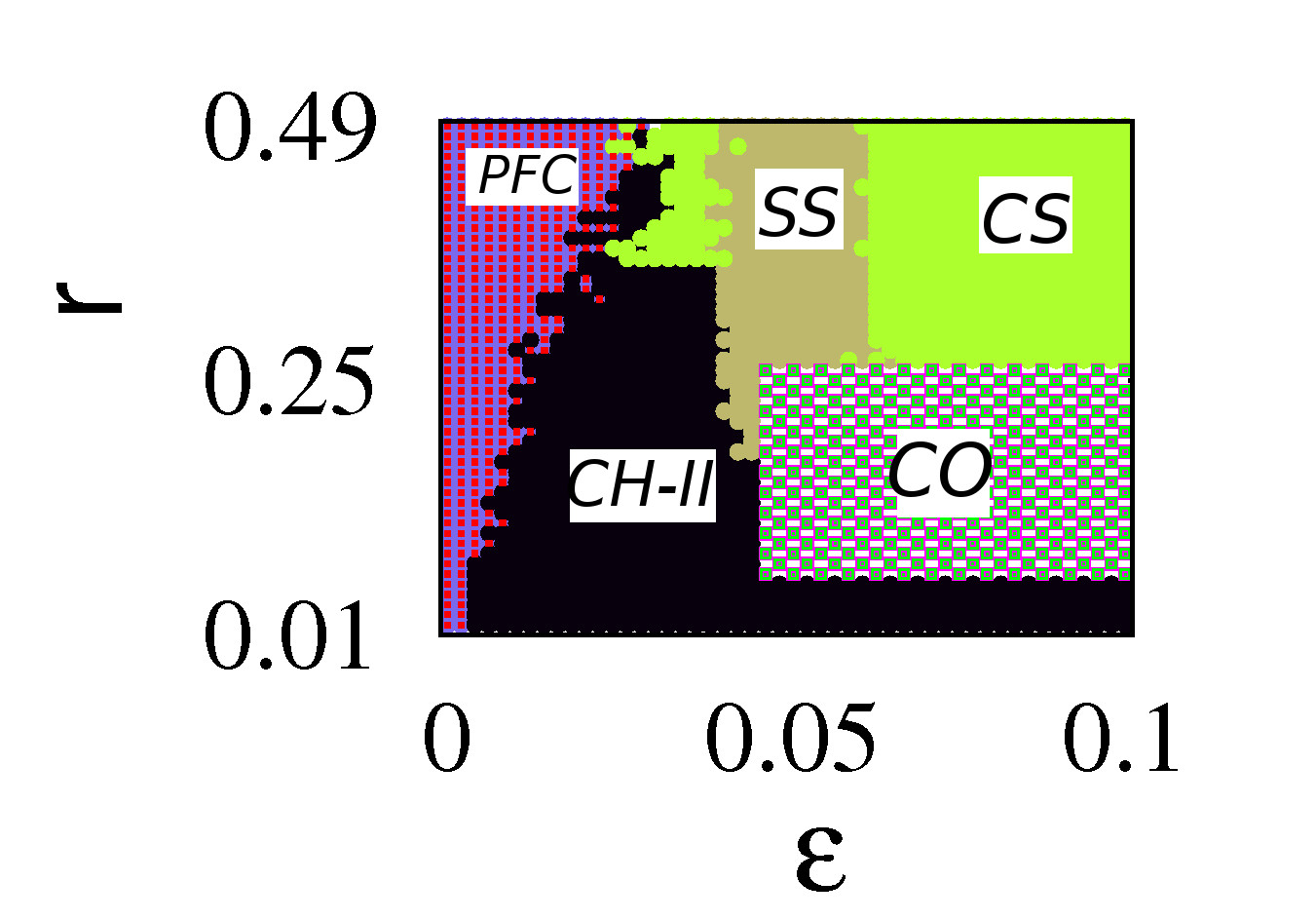}
\caption{(Color online)  (a) Two-phase diagram as a function of the strength of the 
nonlocal coupling $\varepsilon\in(0,0.1)$ and the coupling radius $r\in(0.01,0.49)$ in an array of nonlocally coupled R\"ossler 
oscillators under common  environmental coupling, demarcating different collective dynamical regimes.}
\label{chaos_fig7}
\end{figure}
\begin{figure}
\centering
\includegraphics[width=0.7\columnwidth]{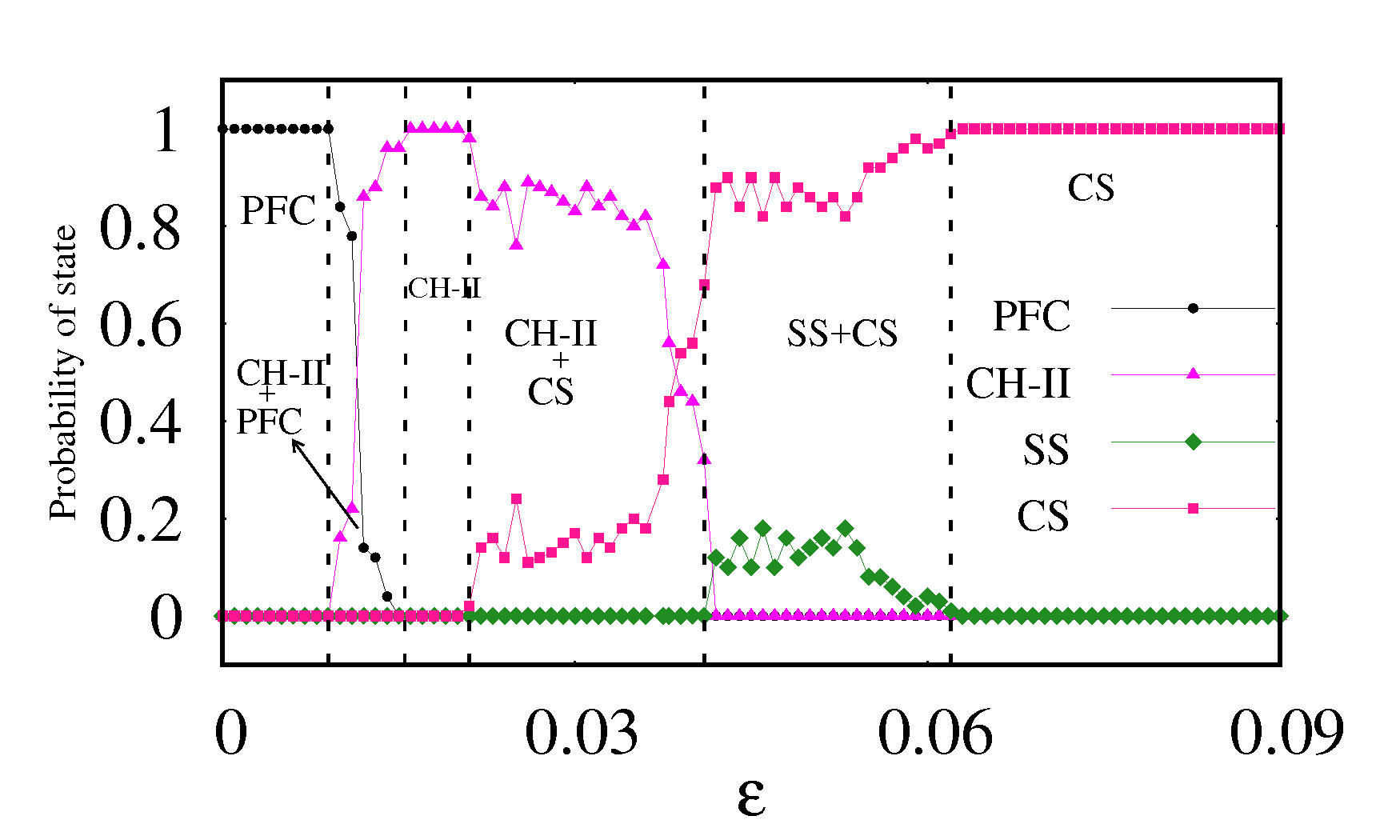}\caption{(Color online) The probability of occurrence of  clusters with phase-flip transition ($PFC$),  chimera at the phase-flip transition ($CH-II$), solitary state ($SS$) and 
complete synchronized state ($CS$)  out of  a set of $100$ different realizations  using $100$ different initial conditions for each oscillators as a function of the 
 nonlocal coupling strength $\varepsilon$ for $q=0.4$, $r=0.3$ and $N=100$.}
\label{multi}
\end{figure}
\begin{figure*}
\centering
\includegraphics[width=1.05\columnwidth]{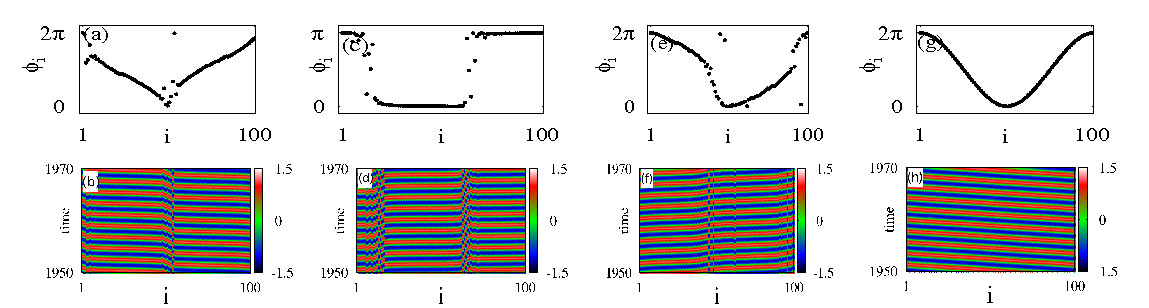}
\caption{(Color online) Snapshots of the instantaneous phases $\phi_{i}$ (top row) and the space-time evolution
(bottom row) of the ensemble of Stuart-Landau oscillators for the coupling radius $r=0.3$ 
and for different values of the strength of the nonlocal coupling (a)-(b) $q=0.4$, $\varepsilon=0.04$,
(c)-(d) $q=0.8$, $\varepsilon=0.04$, (e)-(f) $q=0.8$, $\varepsilon=0.08$, and (g)-(h) $q=0.8$, $\varepsilon=0.12$.
The values of the other parameter are $\omega=10$, $k=25, k_1=0.5, q=0.8$ and $\eta=10$.}
\label{chaos_fig8}
\end{figure*}

We have also quantified the emergent dynamical states in terms of the number of oscillators exhibiting in-phase oscillations,
anti-phase oscillations and desynchronized state as $n_i, n_o$ and $n_d$ to  characterize
the distinct dynamical behaviors observed as a function
of the system parameters. This is  carried out by means of the cross-correlation co-efficient ($CC$) defined as
\begin{equation}
CC_i = \frac{\langle(x_{r}(t)-\langle x_{r}(t)\rangle)(x_{i}(t+\Delta t)-\langle x_{i}(t)\rangle)\rangle_t}{\sqrt{\langle(x_{r}(t)-\langle x_{r}(t)\rangle)^{2}\rangle_t \langle(x_{i}(t)-\langle x_{i}(t)\rangle)^{2}\rangle}_t},
\label{cc}
\end{equation}
where $i=1,2..,N$ and $r\ne i$, $\langle\cdot \rangle_t$ represents the time average, $\Delta t$ is the time shift,
and $x_r$ is the reference oscillator chosen either from the in-phase (complete synchronized/coherent) or anti-phase synchronized
domain.  Thus for in-phase synchronized oscillators the correlation coefficient is characterized by $+1$, whereas for
anti-phase synchronized oscillators $CC_i$ acquires the value $-1$.  The desynchronized oscillators
are characterized by intermediate values between $\pm1$. The number of $+1$, $-1$ and the intermediate values between them
acquired by $CC_i$ provide the estimate of the number of in-phase synchronized oscillators $n_i$, the number of anti-phase synchronized oscillators $n_o$
and the number of desynchronized oscillators $n_d$, respectively.

The numbers of oscillators in each of the above states,  calculated for  $100$ different realizations, are 
shown in  Figs.~\ref{chaos5}(a) and \ref{chaos5}(b) for $q=0.02$ and $q=0.4$, respectively, along with the error bars as  a function 
of the strength of the nonlocal coupling $\varepsilon$. The number of desynchronized oscillators $n_d$ equal 
to the total number of oscillators in the ensemble  in the range of $\varepsilon\in(0,0.008)$ in Fig.~\ref{chaos5}(a) confirms  asynchronous 
evolution of the ensemble of oscillators. The finite number of
in-phase synchronized and desynchronized oscillators in the range of $\varepsilon\in(0.008,0.018)$
elucidates the conventional chimera. For $\varepsilon>0.018$, the number of in-phase synchronized
oscillators equal to the total number of oscillators $N$ corroborates the in-phase synchronous evolution
of the ensemble of oscillators. It is also to be noted that the number of anti-phase synchronized
oscillators remain at null value for the entire range of $\varepsilon$  in Fig.~\ref{chaos5}(a).

In Fig.~\ref{chaos5}(b), the number of oscillators
exhibiting in-phase and anti-phase oscillations are almost equal in number
in the range of $\varepsilon\in(0,0.016)$ corresponding to the existence of $PFC$
without any  oscillators with random phases.  In the range of $\varepsilon\in(0.016,0.041)$,
the number of desynchronized oscillators $n_d$  increases and almost equal to $n_i$ and $n_o$
corroborating the coexistence of anti-phase synchronized coherent domains along with
the desynchronized domains, that is chimera at the phase-flip transition. Beyond 
$\varepsilon=0.041$, there exists almost completely synchronized oscillators as indicated
by $n_i$ close to $N=100$ coexisting along with a few desynchronized oscillators
confirming the existence of solitary states up to $\varepsilon=0.062$. The values of 
$n_i$ attain unity for $\varepsilon>0.062$ elucidating the emergence of completely synchronized state.

 In Fig. \ref{chaos5}(c), 
we have depicted the solitary fraction, namely the ratio of number of desynchronized oscillators $N_{sol}$ to the total number of oscillators $N$
in the ensemble to characterize the solitary state.
We have fixed the value of the coupling strength $\epsilon=0.051$ and  vary the   coupling radius $r$. 
It is evident that the average number of oscillators excursing away from the synchronized state, shown by the line connecting 
filled circles, representing  the solitary state decreases exponentially upon increasing the coupling radius.
The dotted line corresponds to the exponential fit.         
As the coupling radius is increased  to include more number of nearest neighbors, the tendency
of synchronization among them  also increases for a given coupling strength
resulting in more number of synchronized oscillators and a corresponding decrease in
the number of desynchronized oscillators thereby leading to an exponential decay of the 
average number of oscillators excursing away from the synchronized
state. 
It may also be noted that when the coupling radius approaches  the global coupling limit the number of desynchronized 
oscillators (solitary oscillators) in the solitary state decreases while for an intermediate value of the coupling radius ($r=0.25$) the 
solitary oscillators are  large in number.  

 Further, we have also estimated the probability distribution of the correlation coefficient $\rho=pbt(CC_i)$
to characterize the different types of chimeras and solitary state. 
The probability distribution of the correlation coefficient corresponding to the different 
collective dynamical states in the ensemble of R\"ossler oscillators is shown in Fig.~\ref{chaos_fig5_1}.
 The distribution of the correlation
coefficient in Fig.~\ref{chaos_fig5_1}(a) is centered only at $+1$ and $-1$  for  $\varepsilon=0.0$ and $q=0.4$
 elucidating the $PFC$  with only in-phase and anti-phase synchronized states.  The distribution centered about $+1$ 
 with finite distributions between
 $\pm1$ in Fig.~\ref{chaos_fig5_1}(b) for  $\varepsilon=0.01$ and $q=0.02$ characterizes the conventional chimera ($CH-I$)
(see Fig.~\ref{chaos_fig2}(b)) coexisting with in-phase synchronized domain and desynchronized domains. 
The probability distribution localized at both   $+1$ and  $-1$ along with finite distribution between them
in Fig.~\ref{chaos_fig5_1}(c) confirms the chimera at the phase-flip transition ($CH-II$) for
$\varepsilon=0.03$ and $q=0.4$ as it is characterized by out-of-phase synchronized coherent domains
interspersed by asynchronous incoherent domain. The maximal probability at $+1$ with a few intermediate distributions  between
$\pm1$ characterizes the solitary state in Fig.~\ref{chaos_fig5_1}(d) for $\varepsilon=0.051$ and $q=0.4$.

The robustness of the observed dynamical transitions as a function of the strength
of the nonlocal coupling $\varepsilon\in(0,0.1)$ and that of the mean field 
coupling $q\in(0,1)$ is elucidated in the two-phase diagram shown in Fig.~\ref{chaos_fig6}. 
The different dynamical regimes are demarcated using the strength of incoherence $S$, 
number of oscillators exhibiting in-phase oscillations $n_i$,
out-of-phase oscillations $n_o$ and desynchronized state  $n_d$.
The parameter space corresponding to the desynchronized state
is indicated as $IC$ in the range of $\varepsilon\in(0,0.005)$ and  $q\in(0,0.1)$. The parameter
spaces marked as $PFC$, $CH$-$I$ and $CH$-$II$  correspond to the $PFC$, conventional chimera 
and the chimera at the phase-flip transition, respectively. The parameter regime  
corresponding to the phase synchronization is represented by $CO$ while the
parameter spaces indicated by $SS$ and $CS$ correspond to the solitary state 
and completely synchronized state, respectively.  It is also clear from the two-parameter phase diagram
that a set of finite values of  $\varepsilon$ and $q$ is necessary to observe the reported new type of chimera, namely
the chimera at the phase-flip transition.

We have also depicted another two-phase diagram as a 
function of the strength of the nonlocal coupling $\varepsilon\in(0,0.1)$ and
the coupling radius $r\in(0.01,0.49)$ in Fig.~\ref{chaos_fig7}
to illustrate the effect of the coupling radius $r$.  The parameter regimes
corresponding to the  $PFC$ , chimera at the phase-flip transition,
solitary state, in-phase synchronized state (coherent state) and completely synchronized state
are indicated by  $PFC$, $CH$-$II$, $SS, CO$ and $CS$, respectively.  It is also
to be noted that as the radius of the nonlocal coupling increases, the spread of the
$PFC$ also increases as a function of the coupling strength $\varepsilon$.
 On contrary,   smaller values of  coupling radius $r$ favours the emergence of chimera at the phase-flip transition to 
a larger range of $\varepsilon$. It is  evident from Fig.~\ref{chaos_fig7} that as the radius of the nonlocal coupling is decreased,
the spread of the chimera at the phase-flip transition increases as a function of $\varepsilon$.

\subsection{Multistability}
In order to elucidate the multistability of the observed dynamical behaviors, we have estimated the probabilities
of the $PFC$, chimera at the phase-flip transition, solitary state and complete synchronized state  
out of a set of $100$ different realizations using $100$ random initial conditions, where each of the oscillators is distributed between $1$ and $-1$. The parameter
 values are fixed as $q=0.4$ and $r=0.3$ and the probability distribution is depicted in Fig.~\ref{multi} as a 
 function of the strength of the nonlocal coupling.  The dynamical states, namely $PFC$, chimera 
 at the phase-flip transition, solitary state and complete synchronized state, are represented by lines connecting
 filled circles, filled triangles, filled diamonds, and filled squares, respectively.  The probability of the $PFC$
 is unity in the range of $\varepsilon\in(0,0.01)$ confirming that the $PFC$  is the only collective dynamical behavior
 observed in this range of $\varepsilon$.  $PFC$ and chimera at the phase-flip transition coexist 
 in the range of $\varepsilon\in [0.01,0.015)$ illustrating multistability between them. Chimera at the phase-flip transition
 alone exists in the range of $\varepsilon\in [0.015,0.02)$, while it coexists with the complete synchronized state
 in the range of $\varepsilon\in [0.02,0.041)$.  Further, the complete synchronized state coexists with the solitary state
in the range of $\varepsilon\in [0.041,0.064)$. Beyond $\varepsilon=0.064$, one finds that the complete synchronized state is monostable.

\section{Nonlocally coupled Stuart-Landau oscillators with common dynamic environment} 
In this section, we demonstrate the emergence of chimera states at the phase-flip transition with two out-of-phase 
synchronized coherent domains characterizing phase-flip transition
 interspersed by an asynchronous incoherent domain in an  
ensemble of identical Stuart-Landau oscillators with both nonlocal 
and dynamic environmental 
couplings. The corresponding evolution equations are represented as
\begin{subequations}
\begin{align}
\dot{x}_{i}=&\,(1-(x^{2}_{i}+y^{2}_{i}))x_{i}-\omega y_{i}+\frac{\varepsilon}{2P}\sum_{j=i-P}^{j=i+P}(x_{j}-x_{i}),\\ 
\dot{y}_{i}=&\,(1-(x^{2}_{i}+y^{2}_{i}))y_{i}+\omega x_{i}+kw_{i},\\
\dot{w}_{i}=&\,-w_{i}+k_{1}y_{i}-\eta(w_{i}-\frac{q}{N}\sum_{j=1}^{N}w_{j}),\\
& i=1,2,....,N. \nonumber
\end{align}
\label{sl}
\end{subequations}
where $\omega=10$, and $k_{1}=0.5$ are the system parameters. The agents in the common 
environment interact with the oscillators with the  strength  $k=25$ and the diffusion 
coefficient $\eta$ is fixed as $\eta=10$.  In the absence of the nonlocal coupling, all
the oscillators evolve in asynchrony below a threshold value of
the mean field coupling $q$, above which the oscillators in the ensemble evolve in complete
synchrony.  The dynamic environmental coupling is incapable of inducing phase-flip
transition in the ensemble of Stuart-Landau oscillators in the absence of the
nonlocal coupling, whereas the same coupling has been shown
to induce phase-flip transition in two coupled Stuart-Landau oscillators for $\varepsilon=0$~\cite{Amit:12}.
 To examine the collective dynamical behaviors induced by the nonlocal coupling
in the ensemble of Stuart-Landau oscillators coupled indirectly through the
dynamic environment, the strength of the mean field diffusive coupling  $q$ and the  strength of the nonlocal coupling $\epsilon$ 
are varied while  the radius of the nonlocal coupling is fixed as $r=0.3$ and 
the total number of oscillators is fixed as $N=100$. Initial conditions are chosen on the sphere  $x^{2}+y^{2}+w^{2}=1$.

\begin{figure}
\centering
\includegraphics[width=0.6\columnwidth]{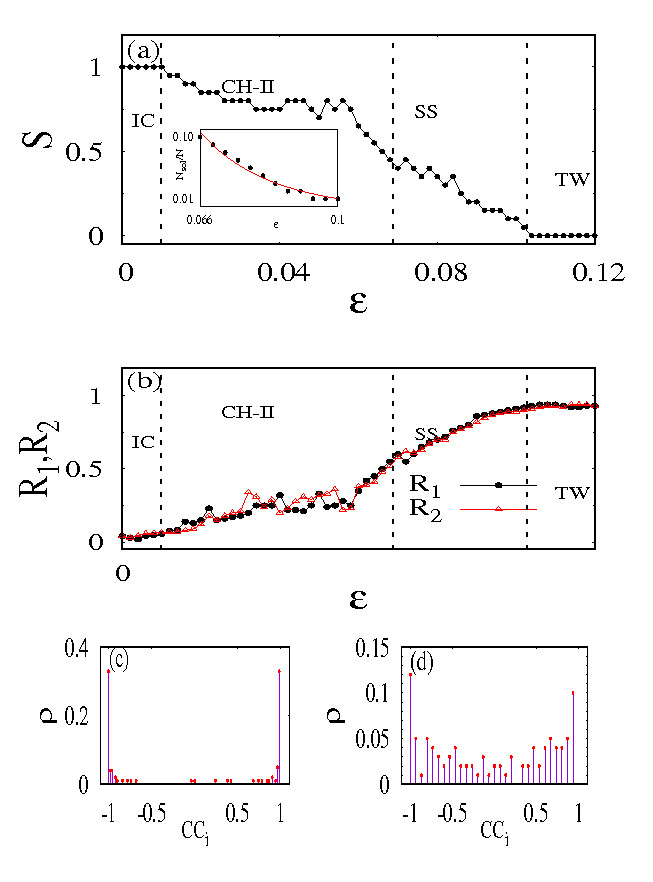}
\caption{(Color online) (a) The strength of incoherence $S$, (b)  Kuramoto order parameter $R_{1}$, $R_{2}$ as a function of the nonlocal coupling strength $\varepsilon$ for  $q=0.8$ (The inset shows that $\frac{N_{sol}}{N}$ versus $\epsilon$, where $N_{sol}$ denotes 
number of oscillators away from synchronized group, namely the solitary fraction). Probability distribution of correlation coefficient for different 
collective dynamical states in the ensembles of Stuart- Landau oscillator 
(c) chimera II ($ q=0.8,\epsilon=0.04$), and
(d) solitary states($q=0.8,\epsilon=0.08$).
}
\label{sl_prob}
\end{figure}
 The snapshots of the instantaneous phases and the spatiotemporal plots of the ensemble
of coupled Stuart-Landau oscillators are depicted in Fig.~\ref{chaos_fig8} for different values of $q$ and $\epsilon$.
The oscillators in the ensemble display conventional chimera  ($CH-I$) for a low value of mean field diffusive
 coupling $q=0.4$ and the nonlocal coupling strength $\epsilon=0.04$ 
(see Figs.~\ref{chaos_fig8}(a) and \ref{chaos_fig8}(b)). Now, the mean field diffusive coupling is increased to $q=0.8$
and by varying the strength of the nonlocal coupling, the ensemble of oscillators is found to split into coherent 
domains exhibiting out-of-phase synchrony interspersed by  an incoherent domain comprised of desynchronized oscillators.
Such a state reveals
the interesting  new type of chimera, namely, chimera at the phase-flip transition. Such a chimera ($CH$-$II$) is shown in 
the snapshot of the instantaneous phases (see Fig.~\ref{chaos_fig8}(c))
of the oscillators and their space-time plot (see Fig.~\ref{chaos_fig8}(d))
for  $\epsilon=0.04$. Further increase in the strength of the nonlocal coupling
leads to the emergence of solitary states where random oscillators hop away
from the synchronized/coherent oscillators as 
illustrated in Figs.~\ref{chaos_fig8}(e) and \ref{chaos_fig8}(f) for $\epsilon=0.08$.
Finally, a travelling wave pattern emerges from the ensemble of oscillators
for large coupling strengths (see  Figs.~\ref{chaos_fig8}(g) and \ref{chaos_fig8}(h) for  $\epsilon=0.12$). 

The strength of incoherence $S$  is depicted in Fig.~\ref{sl_prob}(a) to characterize the observed collective
dynamical behavior in Fig.~\ref{chaos_fig8}.  We have estimated the strength of incoherence for $q=0.8$ and 
hence one can observe only the chimera at the phase-flip transition, solitary state and complete synchronized state
as a function of $\varepsilon$.  The unit value of the strength of incoherence $S$ in the range of  $\varepsilon\in(0,0.01)$
elucidates the asynchronous evolution of the coupled Stuart-Landau oscillators.  Chimera at the phase-flip transition 
is observed in the range of $\varepsilon\in[0.01,0.066)$ as  is indicated by the value of $S$ less than unity 
but greater than $S=0.5$ attributing to the fact that asynchronous oscillators in the incoherent domain of $CH-II$ are
large in number.  Solitary state is observed in the range of $\varepsilon\in[0.066,0.1)$ as the value of $S$ is rather low here
corresponding to a small number of solitary oscillators. The value of the strength of incoherence $S$ decreases smoothly in 
the range attributing to the fact that all the oscillators are in coherent state, and not in the completely synchronized state, other than 
the solitary oscillators.The null value of the  strength of incoherence $S$ 
for $\varepsilon>0.1$ corroborates the  complete synchronized state.  We have also depicted the solitary fraction
of the ensemble of  Stuart-Landau oscillators  to characterize the solitary state 
in the inset of Fig.~\ref{sl_prob}  in the range of  $\varepsilon\in[0.066,0.1)$.   As in the case of Stuart-Landau oscillators, the number of solitary oscillators, shown by line connecting filled circles, decreases exponentially as the nonlocal coupling strength $\varepsilon$ is increased.   The dotted line corresponds to the exponential fit.  It indicates that, as the coupling radius is increased to include more number of nearest neighbors, the tendency of synchronization among them is increased for a given coupling strength resulting in more number of synchronized oscillators and a corresponding decrease in the number of desynchronized oscillators thereby leading to an exponential decay of the  average number of oscillators excursing away from the synchronized state. Similar scaling for solitary states is also reported in Refs.~\cite{jaros2015,kapi,nade2017,prema:2016,brez,ym2014} for a system of nonlocally coupled pendulum like oscillators and Stuart-Landau oscillators. Further, in order to confirm the existence of different collective dynamical states, we also plotted the Kuromoto order parameters $R_{1},R_{2}$ versus coupling strength. It also distinguishes different collective dynamical states in terms of $R_{1}=R_{2}\approx 0$ if phases of oscillators are uniformly distributed, $R_{1}=R_{2}\approx 1$ for travelling wave state, in the case of $CH-II$, the order parameter takes a values of $0\leq R_{1},R_{2}\leq 1$. One may also note that both the strength of incoherence and the Kuramoto order parameter  confirm the transition among the different states is a second order transition as there is a smooth transition from one state to another.

\begin{figure}
\centering
\includegraphics[width=0.7\columnwidth]{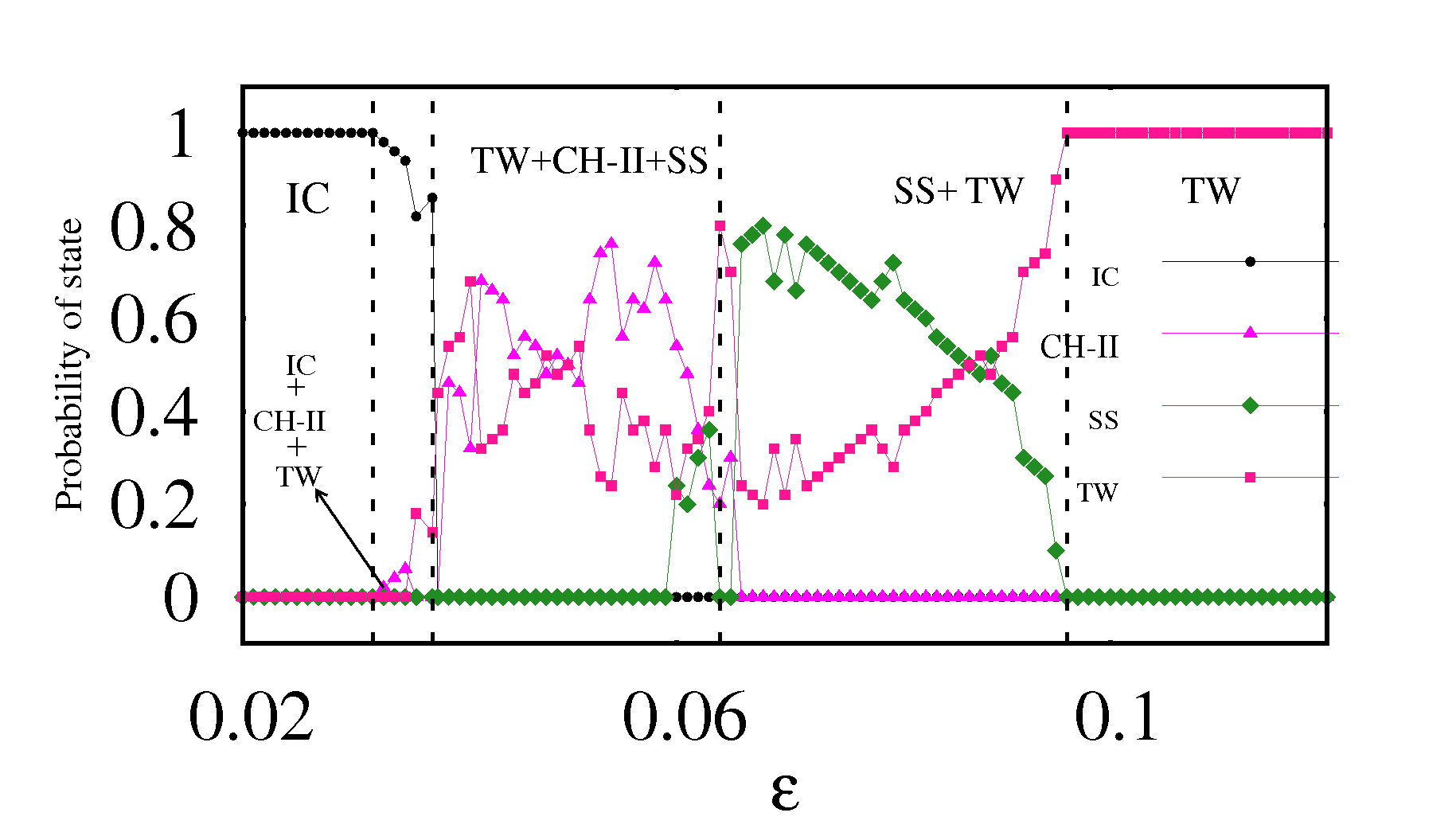}
\caption{(Color online) The probability of  occurrence of asynchronous state ($DSYC$), chimera 
 at the phase-flip transition ($CH-II$) , solitary state ($SS$) and travelling wave pattern ($TW$)  as a function of 
 the nonlocal coupling strength  $\varepsilon$ for $q=0.4$, $r=0.3$ and $N=100$ respectively.}
\label{sl_prob1}
\end{figure}
To characterize the different dynamical behaviors observed in the coupled Stuart-Landau oscillators, we have further depicted the
probability of cross correlation $\rho=CC_i$ in Figs.~\ref{sl_prob}(c) and (d). 
 Now, the value of $q$ is fixed as $q=0.8$
to characterize the $CH-II$  state observed in Fig.~\ref{sl_prob}(a). The probability distributions localized 
at both   $+1$ and  $-1$ along with finite distribution between them in Fig.~\ref{sl_prob}(c) 
confirms the chimera at the phase-flip transition ($CH-II$) for
 $\varepsilon=0.04$.  It is characterized by out-of-phase synchronized coherent domain 
interspersed by an asynchronous incoherent domain. 
The maximal probability at $+1$ with a few intermediate distributions  between
$\pm1$ characterizes the solitary state in Fig.~\ref{sl_prob}(d) for $\varepsilon=0.08$. 

In order to elucidate the multistability nature of the observed dynamical behavior in coupled Stuart-Landau oscillators,
we have estimated the probabilities of the asynchronous state, chimera at the phase-flip transition, solitary state and complete synchronized state   out of $100$ random initial conditions for each oscillators satisfying $x^{2}+y^{2}+w^{2}=1$. The parameter
values are fixed as $q=0.8$ and $r=0.3$ and the probability distribution is depicted in Fig.~\ref{sl_prob1} as a 
function of the strength of the nonlocal coupling.  The dynamical states, namely, asynchronous state, chimera 
at the phase-flip transition, solitary state and travelling wave pattern are represented by line connecting
filled circles, filled triangles, filled diamonds, and filled squares, respectively.  Asynchronous state alone exists 
in the range of $\varepsilon\in(0,0.032)$  as indicated by the null value of the probability distributions of  
various collective states other than that of the asynchronous state.  Chimera at the phase-flip transition ($CH-II$) and  travelling wave ($TW$) 
coexists  along with the asynchronous state in the range of $\varepsilon\in[0.032,0.038)$  attributing to the finite value of their probability distributions. 
$CH-II$ and $TW$ coexist in the range of $\varepsilon\in[0.038,0.06)$ as shown by their probabilities in  
Fig.~\ref{sl_prob1}. Further,  $CH-II$ and $TW$ coexist along with the solitary state ($SS$) in the range of  $\varepsilon\in[0.06,0.066)$, 
whereas $TW$ and $SS$ exists in the range of $\varepsilon\in[0.066,0.097)$ demonstrating their multistable nature.   For $\varepsilon>0.97$
travelling wave pattern alone is stable as indicated by the unit value of its probability.

\begin{figure}
\centering
\includegraphics[width=0.7\columnwidth]{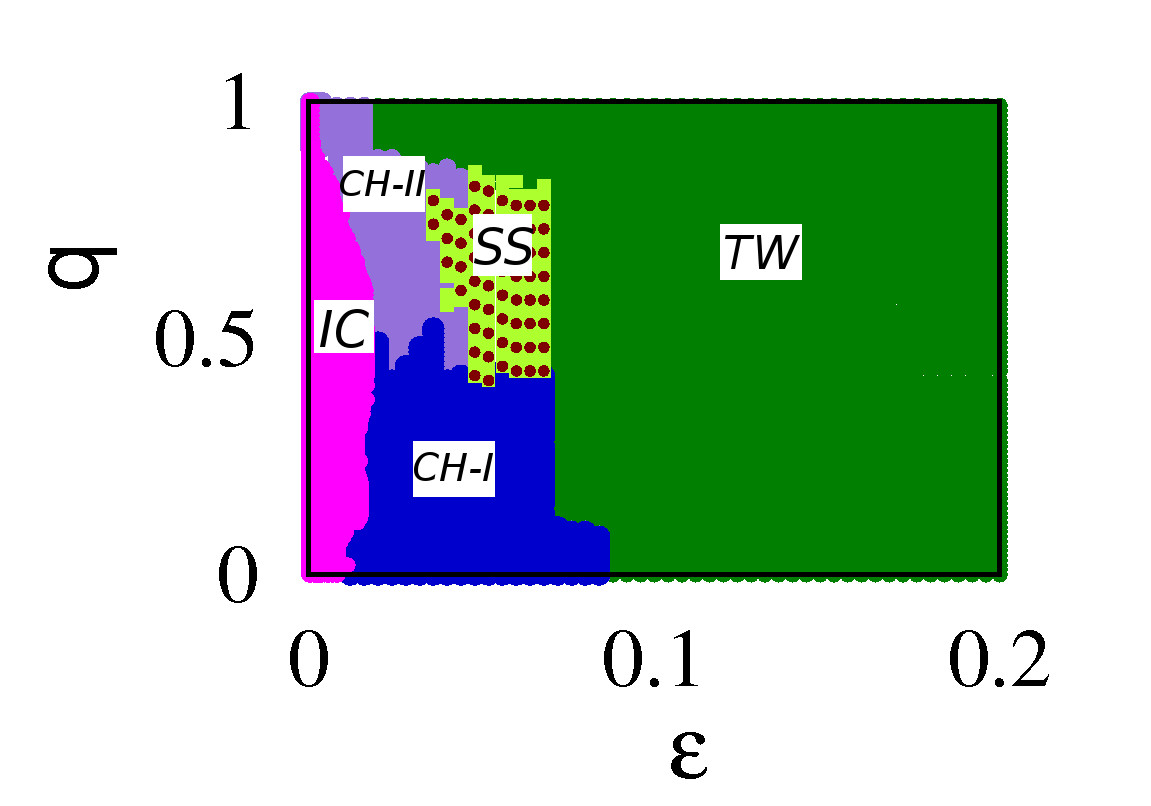}
\caption{(Color online) Two parameter phase diagram depicting the collective dynamical
states of nonlocally coupled Stuart-Landau oscillators with a common dynamic environment as
a function of the strength of the nonlocal coupling $\varepsilon\in(0,0.2)$ and 
the strength of the mean field coupling $q\in(0,1)$. The parameter space marked as
$IC$,  $CH-II$, and $TW$ corresponds to the desynchronized state,
chimera at phase-flip transition, and  travelling wave pattern,
respectively.}
\label{chaos_fig9}
\end{figure}
For a global perspective of the emerging collective dynamics of the ensemble of 
nonlocally coupled Stuart-Landau oscillators with the common environmental coupling, a two-parameter phase
diagram is  depicted in Fig.~\ref{chaos_fig9} as a function of the strength of
the nonlocal coupling $\varepsilon\in(0,1)$ and the strength of the mean field coupling
among the agents $q\in(0,1)$. The parameter space that results in the desynchronized
state is represented by $IC$ in Fig.~\ref{chaos_fig9}.   The set of parameter values 
that can give rise to the chimera state  at the phase-flip transition, solitary state, 
and travelling wave patterns are marked as $CH$-$II$, $SS$, and $TW$, respectively, in the two-phase diagram.

\subsection{Other initial conditions:Stuart-Landau Oscillators}
\begin{figure}
\centering
\includegraphics[width=1.0\columnwidth]{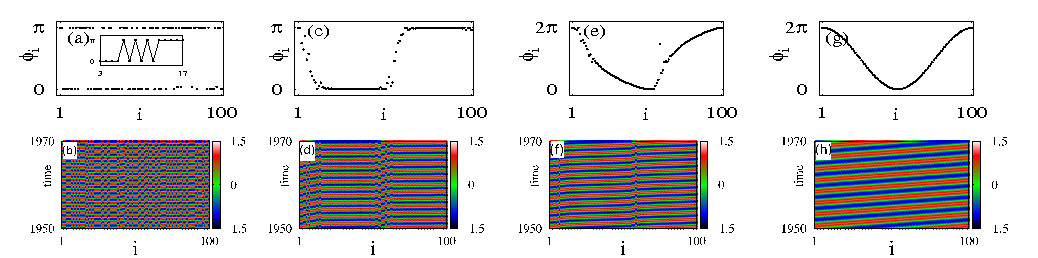}
\caption{(Color online) Snapshots of the instantaneous phases $\phi_{i}$  of the ensemble of Stuart-Landau oscillators for $q=0.8$  
in the value nonlocal coupling (a)-(b) $\epsilon=0.005$ ,(c)-(d)  $\epsilon=0.05$, (e)-(f) $\epsilon=0.13$ and 
(g)-(h) $\epsilon=0.22$ (initial condition in the form of  cluster state). The values of the parameter are the same as 
in Fig.~\ref{chaos_fig8}.}
\label{appen_fig1}
\end{figure}

\begin{figure}
\centering
\includegraphics[width=0.6\columnwidth]{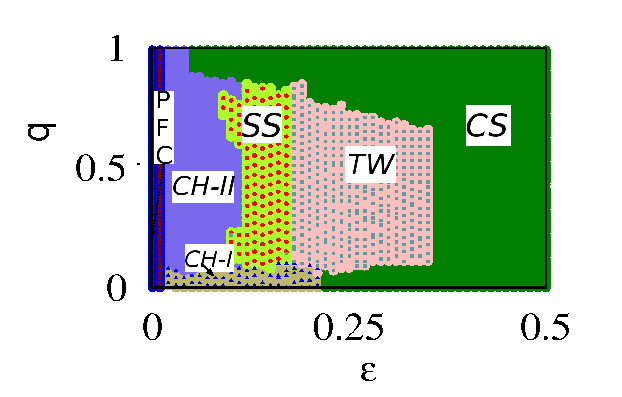}
\caption{(Color online) Two parameter phase diagram as a function of the strength of the 
nonlocal coupling $\varepsilon\in(0,0.5)$ and  the strength of the mean field coupling 
$q\in(0,1)$ depicting different collective dynamical states of the ensemble of identical Stuart-Landau oscillators
for specially prepared initial conditions in the form of two-cluster states. The parameter space marked as
$PFC$, $CH-I$, $CH-II$, $SS$, $TW$ and $CS$ corresponds to the  $PFC$, conventional chimera,
chimera at phase-flip transition,  solitary state,  travelling wave pattern, and complete synchronized state,
respectively.}
\label{appen_fig2}
\end{figure}

Further, it is to be noted that the emergence of   clusters with phase-flip transition ($PFC$) or chimera at the phase flip transition state highly relies on the nature of distribution of the initial conditions of the ensemble of oscillators. As we have already pointed out, uniform random numbers between $-1$ and $1$ is chosen as initial conditions to obtain the discussed dynamical behaviors of R\"ossler  oscillators.  $PFC$ states are not possible in the ensemble of Stuart-Landau oscillators for the initial conditions distributed between $-1$ and $1$. On the other hand, if we prepare the initial conditions in the form of two cluster states (that is, half of the oscillators are provided $+1$ as initial conditions and the other half as $-1$ corresponding to the in-phase and out-of-phase states of $PFC$,  the  $PFC$ states (two-cluster state)  emerge even in the absence of nonlocal coupling, but only with the environmental coupling. However, the $PFC$ shown in Figs.~\ref{appen_fig1}(a)-(b) is plotted for a finite value of nonlocal coupling $\varepsilon=0.005$ to break the permutation symmetry
 facilitated by the mean field global coupling, so that the state depicted in Figs.~\ref{appen_fig1}(a)-(b) cannot be simply regarded as.a two cluster state.  The values of the other parameters are the same as in Fig.~\ref{chaos_fig8}. Upon increasing the strength of the nonlocal coupling $\varepsilon$, we have observed the chimera at the phase-flip transition (see Figs.~\ref{appen_fig1}(c)-(d)), solitary states (see Figs.~\ref{appen_fig1}(e)-(f)) and travelling wave pattern (see Figs.~\ref{appen_fig1}(g)-(h)).  We have also depicted the variety of collective dynamical behaviors exhibited by the ensemble of identical Stuart-Landau oscillators in the two-parameter phase diagram (see Fig.~\ref{appen_fig2}) for the above choice of initial conditions in the form of two-cluster states.


\section{Conclusion}
We have demonstrated the emergence of an interesting new type of chimera, where
the two adjacent coherent domains exhibit out-of-phase synchronized oscillations
similar to phase-flip transitions.  The incoherent domain of the chimera 
is constituted by the asynchronous oscillators at the onset of the phase-flip transition
among the coherent domains, where each domain exhibits phase synchronized oscillations.
Such a set of coexisting   out-of-phase synchronized adjacent coherent domains
interspersed by asynchronous incoherent domains constitute a rich spatio-temporal behavior constituting a new
type of chimera, namely chimera at the phase-flip transition. Such a dynamical behavior is
observed only in the nonlocally coupled ensemble of identical oscillators with
a common dynamic environmental coupling. The nonlocal coupling facilitates
the emergence of the asynchronous incoherent domain at the onset of
the phase-flip transition among the ensemble of the identical oscillators. 
The dynamic environment coupling facilitates the onset of phase-flip transition 
in the ensemble of  R\"ossler oscillators,  while both the environmental and nonlocal couplings are
required for the onset of phase-flip transition in the ensemble of Stuart-Landau oscillators. 
In the absence of the nonlocal coupling phase-flip transition will not be
observed  in the ensemble of identical Stuart-Landau oscillators.  We have demonstrated the
generic nature of our results using the paradigmatic R\"ossler and Stuart-Landau oscillators.
In addition, a range of collective behaviors including  clusters with phase-flip transition ($PFC$), conventional chimera,
solitary states,  coherent states, complete synchronization and travelling wave patterns are observed during the dynamical transitions as a function
of the system parameters.

\section*{Acknowledgments}
The work of VKC forms part of a research project sponsored by INSA Young Scientist Project under Grant No. SP/YSP/96/2014. 
DVS is supported by the SERB-DST Fast Track scheme for young
scientist under Grant No. ST/FTP/PS-119/2013 and CSIR Grant No. 03(1400)/17/EMR-II. ML  is supported by a NASI Platinum Jubilee Senior Scientist Fellowship. He is also supported by a CSIR research project and a DST-SERB research project.  \\

\begin{appendix}

\section*{Appendix A: Clusters with phase-flip transition ($PFC$) for other initial conditions} 

\begin{figure}
\centering
\includegraphics[width=1.00\columnwidth]{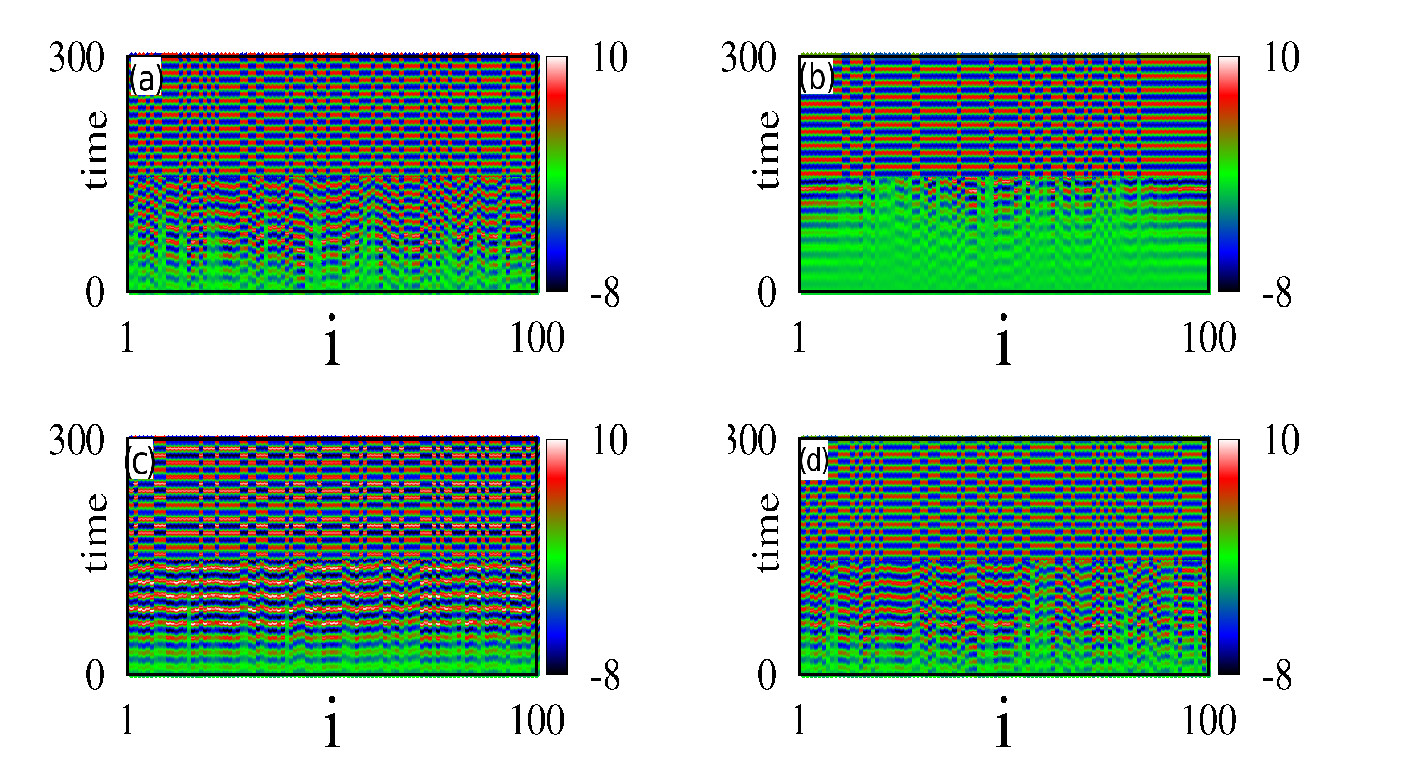}
\caption{(Color online)   Clusters with phase-flip transition ($PFC$)   with transient behavior for further different sets of initial conditions (a) $x^2+y^2+z^2=1$,(b)$sin(\frac{\pi i}{N})$, 
(c) symmetric distribution of uniform random numbers between $p=q=0.5$ and (d) asymmetric distribution of random numbers between $p=0.5$, $q=1.5$.
 Other parameters are the same as in Fig 1(c)-(d).}
\label{initial_condi}
\end{figure}

In order to confirm the robustness of  clusters with phase-flip transition, they are shown using different sets of initial conditions.   $PFC$ emerging out of further two different sets of specific initial conditions $x^2+y^2+z^2=1$ and $\sin(\frac{\pi i}{N})$, where $i=1,2,3....N$, are depicted along with the transients  in Figs \ref{initial_condi}(a) and (b), respectively.  Further, we have also corroborated the robustness of the  clusters with phase-flip transition by showing its existence for two more random initial conditions in Figs \ref{initial_condi} (c) and (d).  
The initial conditions are  uniform random numbers distributed between   $-p$ and $q$.  Random numbers distributed symmetrically between  $p=q=0.5$ are used to show  $PFC$ in Fig\ref{initial_condi} (c) and whereas asymmetric distribution of random numbers between $p=0.5$, $q=1.5$ are used to depict the   $PFC$ in Fig\ref{initial_condi} (d).
These figures elucidate that such a  cluster with phase-flip transition can always emerge in the presence of the dynamic environmental coupling and that they arise  not due to the manifestation of a specific choice of initial conditions.

\section*{Appendix B: Phase flip transition between two coupled R\"ossler  oscillators}

\begin{figure}
\centering
\includegraphics[width=0.5\columnwidth]{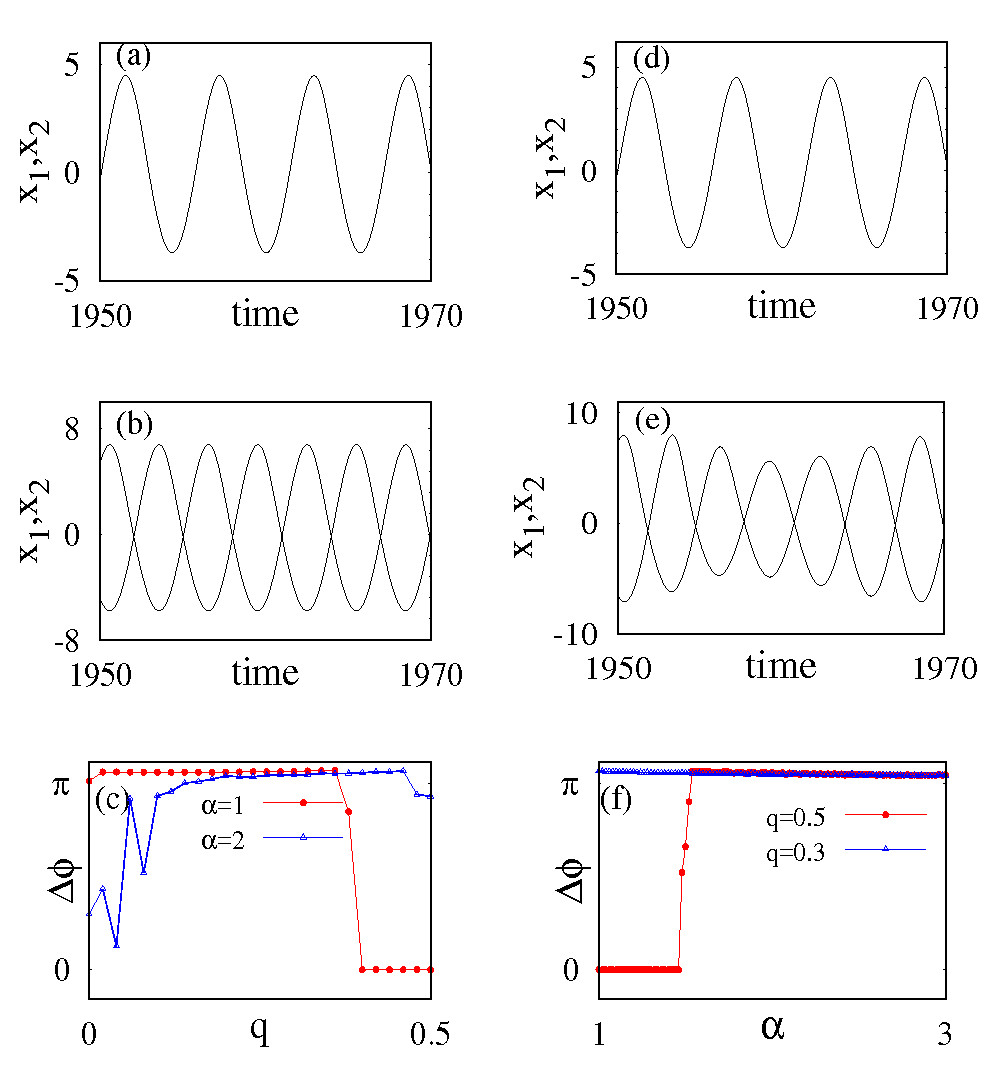}
\caption{(Color online) In-phase (a) $q=0.5$,$\alpha=1$, (d) $\alpha=1$, $q=0.4$ and out-of-phase transitions (b) $q=0.25$, $\alpha=1$, (e) $\alpha=2$, $q=0.4$. The phase difference between the two oscillators as a function of (c) $q$ and (f) $\alpha$ respectively.}
\label{appen_fig3}
\end{figure}

In order to examine the existence of  phase-flip transition as reported in Ref~\cite{Amit:12}, we also  consider two coupled chaotic R\"ossler oscillators interacting through dynamic environment, 

\begin{align*}
\dot{x}_{i}=&\,-y_{i}-z_{i}+\varepsilon(x_j-x_i),  \\  \nonumber
\dot{y}_{i}=&\,x_{i}+ay_{i}, \\   \nonumber
\dot{z}_{i}=&\,b+z_{i}(x_{i}-c)+kw_{i}, \\  \nonumber
\dot{w}_{i}=&\,-\alpha w_{i}+0.5z_{i}+\eta(q\bar{w}-w_{i}), \\  \nonumber
\end{align*} 

where $i, j=1,2$ and $i\ne j$. The parameters are $\bar{w}=\sum_{i=1}^{2}\frac{1}{2}w_{i}$, $a=0.165$, $b=0.4$ and $c=8.5$  are the system parameters and $\eta=2$, $k=24$. We show that the results of in-phase and anti-phase synchronized states for  the values of $q$ and $\alpha$ as shown in Figs~\ref{appen_fig3}(a)-(b) and \ref{appen_fig3}(d)-(e) respectively.  Therefore, the transition between the in-phase and out-of-phase synchronizations is studied numerically using the values of average phase-difference between the coupled oscillator systems. In this case,  the instantaneous phase $\phi_{i}$ of the $i^{th}$ oscillator is defined as $\phi_i=\arctan(y_i/x_i)$, where $x_{i}$ and $y_{i}$ denote the state variables. The average phase difference  between  any two oscillators is  $\Delta \phi_{ij}=<|\phi_{i}-\phi_{j}|>_{t}$. The corresponding results of phase difference between the oscillators  are also plotted as a function of $q$ and $\alpha$ as shown in Fig.~\ref{appen_fig3}(c) and (f), respectively.

\section*{Appendix C: Bistability between two coupled R\"ossler  oscillators}
\begin{figure}
\centering
\includegraphics[width=0.9\columnwidth]{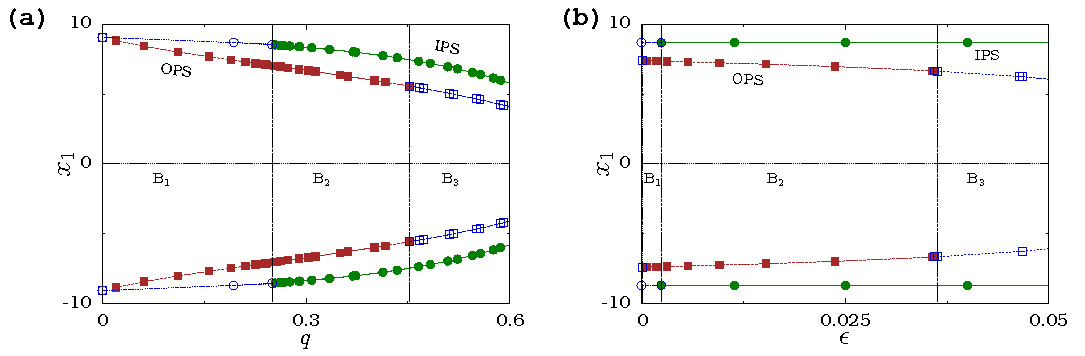}
\caption{(Color online) Two coupled chaotic R\"ossler oscillators interacting through dynamic environment. Bifurcation diagram for varying the values of $q$ (left) for fixed value of coupling strength $\varepsilon=0$ and varying the values of $\epsilon$ (right) for fixed value of $q=0.2$. The unfilled symbols correspond to unstable states, while the filled
symbols represent stable states. The filled circle and filled square represent the in-phase and out-of-phase synchronized states, respectively.}
\label{appen_fig4}
\end{figure}

Now, we will demonstrate the emergence of  bistability  and the related bifurcations/transitions even between a system of two coupled R\"ossler oscillators among the  in-phase 
and out-of-phase synchronized oscillations in order to substantiate our discussion on the observed multistability among $N=100$ oscillators. 
The bifurcation diagrams as  functions of both the strength of the mean field interaction $q$ and the strength of the nonlocal coupling $\varepsilon$ are 
shown in Figs.~\ref{appen_fig4}(a) and (b), respectively.  Figure~\ref{appen_fig4}(a) is depicted  in the absence of any nonlocal coupling,
 $\varepsilon=0$, while Fig.~\ref{appen_fig4}(b) is depicted for $q=0.2$, where there is no bistability in the absence of nonlocal coupling
 as is evident from Fig.~\ref{appen_fig4}(a),  in order to demonstrate that both the mean field and nonlocal couplings
 can induce bistability.   Out-of-phase synchronized oscillations (represented by filled squares) are only stable in $B_1$ and 
 in-phase synchronized oscillations (represented by filled circles) are alone
 stable in $B_3$ in both the figures, whereas in $B_2$ both the out-of-phase synchronized and in-phase synchronized oscillations
 are stable thereby elucidating the emergence of bistability between two different states even in just two coupled oscillators. Therefore
intuitively  one  will be able to realize that for $N=100$ coupled oscillators there exists a rich variety of multistablility among the synchronized and
desynchronized states thereby leading to the chimera states and solitary states discussed in Fig.~\ref{multi}.  Transitions among the
different states are better visualized for $N=100$ coupled oscillators as discussed in the main text.

\end{appendix}

\end{document}